\documentclass[aps,twocolumn,pra,superscriptaddress,amsmath,amssymb]{revtex4-2} 
\usepackage{dcolumn}
\usepackage{amsmath}
\usepackage{mathrsfs}
\usepackage{txfonts}
\usepackage{bm}
\usepackage[T1]{fontenc}
\usepackage{xspace}
\usepackage{ulem}
\usepackage{comment}
\usepackage{braket}
\usepackage{lipsum}
\usepackage{bbold}
\usepackage{mathtools}
\newcommand{\Bigbraket}[1]{\Big\langle #1 \Big\rangle}

\ifx\pdfoutput\undefined
\usepackage[dvipdfmx]{graphicx}
\usepackage[dvipdfmx]{hyperref}
\usepackage[dvipdfmx]{color}
\usepackage[dvipdfmx]{xcolor}
\else
\usepackage{graphicx}
\usepackage{hyperref}
\usepackage{color}
\usepackage{xcolor}
\usepackage[version=3]{mhchem}
\fi
\hypersetup{
        colorlinks=true,
        citecolor=blue,
        urlcolor=blue,
        linkcolor=blue
}

\begin{document}
\let\emph\textit

\title{
    Schwinger boson theory for $S=1$ Kitaev quantum spin liquids
}
\author{Daiki Sasamoto}
\email[sasamoto.daiki.r6@dc.tohoku.ac.jp]{}
\author{Joji Nasu}
\affiliation{
  Department of Physics, Graduate School of Science, Tohoku University, Sendai, Miyagi 980-8578, Japan
}

\date{\today}
\begin{abstract}
The Kitaev model is an exactly solvable model that exhibits a quantum spin liquid as its ground state.
While this model was originally proposed as an $S=1/2$ spin model on a honeycomb lattice, extensions to higher-spin systems have recently attracted attention.
In contrast to the $S=1/2$ case, such higher-$S$ models are not exactly solvable and remain poorly understood, particularly for spin excitations at finite temperatures.
Here, we focus on the $S=1$ Kitaev model, which has been proposed to host bosonic quasiparticles.
We investigate this model using Schwinger boson mean-field theory, where bosonic spinons are introduced as fractional quasiparticles by extending bond operators to address anisotropic spin interactions.
We determine the mean-field parameters that realize a quantum spin liquid in both ferromagnetic and antiferromagnetic Kitaev models.
Based on this mean-field ansatz, we calculate the dynamical and equal-time spin structure factors.
We find that, when one uses the conventional decoupling scheme based on Wick decoupling with respect to spinons to calculate spin correlations, the resultant spin structure factors exhibit a momentum dependence that is not consistent with the sign structure expected from the exchange interaction:
they possess a strong spectral weight indicating ferromagnetic (antiferromagnetic) correlations in the antiferromagnetic (ferromagnetic) Kitaev model.
To resolve this issue, we propose an alternative scheme for evaluating the spin correlations, which is based on decoupling with respect to the bond operators.
We demonstrate that, in our scheme, this discrepancy is removed, and the momentum dependence of the spin structure factors is consistent with the sign of the exchange constant.
We also calculate the temperature evolution of the dynamical spin structure factor and find that the continuum observed at zero temperature splits into two distinct structures as the temperature increases, which can be understood in terms of the bandwidth narrowing of spinons.
Finally, we clarify the origin of why the two distinct decoupling schemes result in different momentum dependences of the spin structure factors and discuss their relationship to results obtained in previous studies.
\end{abstract}
\maketitle


\section{Introduction}
\label{introduction}
Quantum spin liquids (QSLs) are quantum states of insulating magnets that lack long-range order even at zero temperature due to strong quantum fluctuations~\cite{Anderson-1973,Balents-2010}.
They realize highly entangled many-body states that cannot be captured within a classical spin description and have been a central topic in modern condensed matter physics~\cite{
Ramirez-1994,
Balents-2010,
Savary-Balents-2016,
Zhou-Kanoda-2017,
Knolle-Moessner-2019,Wen-Yu-2019,
Broholm-Cava-2020,
Clark-Abdeldaim-2021}.
From a theoretical standpoint, QSLs are a fascinating subject, as they host quasiparticle excitations fractionalized from spin degrees of freedom and exhibit nontrivial topological properties in the ground state, which lie outside the traditional Landau symmetry-breaking paradigm.
However, the theoretical understanding of QSLs remains limited because one of the key ingredients for their realization is strong spin frustration, which hinders the application of conventional analytical and numerical methods.

In this context, the proposal of the Kitaev model profoundly reshaped research on QSLs~\cite{Kitaev-2006}.
This is an $S=1/2$ quantum spin model defined on a honeycomb lattice, offering a rare exactly solvable instance that realizes a QSL ground state in two dimensions.
The realization of the QSL ground state is supported by the presence of a local conserved quantity defined on each hexagonal plaquette, enabling a mapping of the original spin model to a system of free Majorana fermions coupled to static $Z_{2}$ gauge fields.
This mapping implies that the elementary excitations from the QSL ground state consist of Majorana fermions and $Z_{2}$ gauge fluxes (visons), as a manifestation of spin fractionalization.
When time-reversal symmetry is broken by a weak  magnetic field, each vison binds a Majorana zero mode, which behaves as a non-Abelian anyon.
Such a composite quasiparticle is potentially applicable to fault-tolerant topological quantum computation~\cite{Kitaev-2003}.
Importantly, the Kitaev model is not merely a toy model but is relevant to real materials; it has been proposed to be realized in various transition metal compounds with strong spin-orbit coupling, such as iridates and $\alpha$-RuCl$_{3}$~\cite{
Jackeli-Khauliullin-2009,
Chaloupka-2010,Singh-Gegenwart-2010,
Comin-Levy-2012,
Sohn-2013,Chaloupka-2013,Foyevtsova-2013,
Plumb-2014,Katukuri-2014,Yamaji-2014,
Chun-2015,Kubota-2015,
Yadav-2016,Winter-2016,Sinn-2016,
Winter-2017,
Haraguchi-2018,
Takagi-2019,
Haraguchi-2020,Motome-2020,
Jang-2021}, 
and manifestations of spin fractionalization have been observed in such candidate materials~\cite{
Rau-Lee-Kee-2016,
Hermanns-Kimchi-2018,
Motome-Nasu-2020,
Trebst-Hickey-2022,
Rousochatzakis-Perkins-Luo-Kee-2024}.

Recently, the Kitaev model has been generalized to higher-spin systems.
While a local conserved quantity exists on each hexagonal plaquette even in higher-spin systems~\cite{Baskaran-Sen-Shankar-2008,Ma-2023}, the remaining degrees of freedom cannot be mapped onto free quasiparticle systems except in the $S=1/2$ case.
Recently, candidate materials for realizing higher-spin Kitaev models have been proposed, and both experimental and theoretical efforts have been devoted to investigating candidate Kitaev magnets with effective spin $S>1/2$
~\cite{
Baskaran-Sen-Shankar-2008,
Koga-2018,Xu-2018,Oitmaa-2018,Suzuki-2018,
Kim-2019,Stavropoulos-Kee-2019,Minakawa-2019,
Koga-2020,Hickey-2020,Xu-2020,Lee-Kawashima-Kim-2020,Lee-2020,Dong-Sheng-2020,Zhu-2020,
Samarakoon-2021,Lee-Suzuki-Kim-Kawashima-2021,Khait-Stavropoulos-2021,Stavropoulos-2021,
Jin-2022,Fukui-Kato-Nasu-Motome-2022,Bradley-2022,Chen-2022,
Ma-2023,Taddei-2023,
Georgiou-2024}.
For example, a pseudofermion functional renormalization group (pf-FRG) study of the Kitaev-Heisenberg model indicates that a QSL ground state remains stable in the $S=1$ Kitaev model~\cite{Fukui-Kato-Nasu-Motome-2022}.
More interestingly, Ref.~\cite{Ma-2023} introduced a Majorana-parton construction for general spin $S$, in which each spin operator is represented by $8S$ Majorana fermions.
Within this formulation, the extensive plaquette integrals of motion are exactly identified as $\mathbb{Z}_2$ gauge flux operators, revealing an emergent $\mathbb{Z}_2$ gauge structure in the higher-spin Kitaev model despite the absence of exact solvability.
Furthermore, the corresponding $\mathbb{Z}_2$ gauge charges are shown to be composite excitations whose statistics exhibit an even-odd effect: they are fermionic for half-integer spins but bosonic for integer spins.
This result is consistent with the analysis in the anisotropic limit of the Kitaev model, where the system reduces to a set of isolated dimers~\cite{Minakawa-2019}.
Note that the Kitaev interaction in $S = 1$ systems is derived as an antiferromagnetic coupling through perturbation expansions from the strong correlation limit in consideration of realistic materials~\cite{Stavropoulos-Kee-2019}.
Motivated by these findings, the $S=1$ antiferromagnetic Kitaev model has been studied using a Schwinger boson mean-field theory (SBMFT)~\cite{Ralko-Merino-2024}.

The Schwinger boson approach is a powerful analytical method for investigating QSLs in frustrated magnets~\cite{
Arovas-Auerbach-1998,
Read-Sachdev-1991,Sachdev-Read-1991,
Sachdev-1992}.
In this framework, spin operators are represented as products of two bosonic quasiparticles called spinons, and spin interactions are reformulated as effective interactions between these spinons.
The simplest way to treat the resulting interacting boson model is to apply a mean-field approximation, which maps the original spin model onto a free-boson model coupled to self-consistent gauge fields in order to incorporate intersite singlet correlations.
This procedure is called SBMFT.
Note that the mean-field decoupling is performed in terms of bond operators defined on the links of the lattice.
Thus far, SBMFT has been widely employed to investigate QSLs in various frustrated magnets.  
In particular, this approach provides a unified and versatile framework for describing QSLs within a controlled mean-field setting and successfully captures the properties of the ground state and excitation spectra in the antiferromagnetic Heisenberg model on frustrated lattices, such as the triangular lattice~\cite{
Gazza-1993,
Lefmann-Hedegard-1994,
Mattsson-1995,
Misguich-1998,
Yoshioka-1991,
Shen-Zhang-2002,
Lauchli-2006,
Li-2009,Isakov-2009,
Mezio-2011,Feng-2011,
Merino-Holt-Powell-2014,
Lima-2016-4,
Pires-2017,Bauer-2017,Lima-2017-2,Kos-Punk-2017,
Wu-2019,
Zhang-Li-2021} and the kagome lattice~\cite{
Manuel-Trumper-1994,
Li-Su-Shen-2007,
Messio-Cepas-2010,
Fak-2012,Messio-Bernu-2012,
Messio-Lhuillier-2013,
Halimeh-Punk-2016,
Mondal-2017,Chern-Hwang-2017,Messio-Bieri-2017,
Ghosh-Kumar-2018,
Halimeh-Singh-2019,Mondal-2019,
Mondal-2021,
Li-Li-2022,Lugan-Jaubert-2022,
Rossi-Motruk-2023}. 
This framework has also been applied to quantum spin models on the honeycomb lattice, where frustration originates from further-neighbor Heisenberg couplings~\cite{
Mattsson-Frojdh-1994,
Wang-2010,
Cabra-Lamas-2011,
Vaezi-Mashkoori-2012,
Moura-Pereira-2013,Zhang-Lamas-2013,
Zhang-Arlego-2014,Arlego-Lamas-2014,
Pires-2015-2,Pires-2015,
Pires-2016,Lima-2016,
Lima-2017,Jia-2017,
Zhang-Lamas-2018,Merino-Ralko-2018,
Lima-2021,
Lima-2022}.  
By mapping interacting spins onto free bosons coupled to a self-consistent gauge field, SBMFT allows one to evaluate the stability of a QSL ground state, while the instability toward magnetic order can be described as the Bose-Einstein condensation of these bosons at the specific momentum corresponding to the ordering pattern.  
Moreover, SBMFT provides a systematic framework to study excitation spectra at finite temperatures, thereby enabling the exploration of not only ground-state properties but also spin dynamics in a consistent manner.  
SBMFT has been applied to various QSLs.  
For instance, chiral spin liquids, time-reversal-symmetry-breaking QSLs proposed by Kalmeyer and Laughlin~\cite{Kalmeyer-Laughlin-1987}, can be treated within this framework by imposing a time-reversal-symmetry-breaking mean-field ansatz.  
It has been suggested that such ansatz may be stabilized in kagome lattice~\cite{
Messio-Bernu-2012,
Messio-Lhuillier-2013,
Halimeh-Punk-2016,
Messio-Bieri-2017,
Mondal-2017} and honeycomb lattice~\cite{Wang-2010,Ralko-Merino-2024}, which has stimulated the exploration of chiral spin liquids in real materials.  
It is known that various types of QSLs can be systematically classified within the framework of the projective symmetry group (PSG)~\cite{Wen-1989,Wen-2002}, which has been extended to bosonic partons~\cite{Wang-Vishwanath-2006} and widely used in the context of the Schwinger boson approach~\cite{
Hwang-Dodds-2013,Messio-Lhuillier-2013,
Xu-2016,
Kos-Punk-2017,Chern-Hwang-2017,
Li-2019,
Schneider-2022,
Rossi-Motruk-2023}.

As mentioned earlier, the $S=1$ Kitaev model has been studied using SBMFT~\cite{Ralko-Merino-2024}.
That study demonstrated that a QSL ground state remains stable even in the $S=1$ case by analyzing the antiferromagnetic Kitaev model.
It also highlighted the possibility of a chiral spin-liquid state as a viable mean-field ansatz and emphasized the relevance of time-reversal-symmetry-breaking mean-field states based on the zero-temperature dynamical spin structure factor.
At the same time, some time-reversal-symmetric ansatze were found to exhibit features that appeared inconsistent with antiferromagnetic correlations, such as features indicative of ferromagnetic correlations in the spin structure factor.
However, it was not conclusively established whether this behavior should be attributed to the mean-field ansatz itself or to the specific scheme used to evaluate spin correlation functions within SBMFT.
This observation motivates the present study, in which we focus on elucidating the spin dynamics and finite-temperature properties of the $S=1$ Kitaev quantum spin liquid within a given mean-field description.
In this context, a careful treatment of spin correlation functions is essential for a reliable interpretation of the dynamical response.

In this paper, we investigate the $S=1$ Kitaev model within the SBMFT framework by extending the bond-operator representation to incorporate Ising-type interactions.
We introduce a new set of bond operators~\cite{
Kargarian-Langari-Fiete-2012,
Kos-Punk-2017,
Samajdar-Scheurer-2019,
Mondal-2021,
Schneider-2022,
Ralko-Merino-2024}, in addition to the conventional SU($2$)-invariant ones, thereby enabling a unified description of arbitrary-spin interactions.
We reformulate SBMFT using these bond operators and derive the corresponding mean-field Hamiltonian.
We apply this framework to both antiferromagnetic and ferromagnetic $S=1$ Kitaev models on the honeycomb lattice and determine the mean-field parameters self-consistently.
We find that the QSL ground state is stable in the $S=1$ case, but the spinon gap is considerably smaller than that in the $S=1/2$ case.
This gap closes slightly above $S=1$ as the spin length $S$ increases.
The small spinon gap manifests itself in the low-energy spin-excitation spectrum.
To compute the dynamical spin structure factor, we introduce two distinct evaluation schemes for the spin-spin correlator: one is the conventional Wick decomposition with respect to the spinon operators, and the other is a decoupling formulated in terms of the bond operators.
We show that the former leads to a momentum dependence that is not consistent with the antiferromagnetic exchange sign, in which a low-energy structure indicative of ferromagnetic correlations emerges despite the antiferromagnetic Kitaev interaction.
By contrast, the latter yields a momentum dependence more directly consistent with antiferromagnetic correlations. 
In this context, we reexamine a time-reversal-symmetric spin-triplet ansatz, complementary to the time-reversal-symmetry-breaking spin-liquid state proposed as a promising candidate in previous studies, as a representative mean-field state for studying spin dynamics and the role of the correlation-function decoupling scheme.
Motivated by this observation, we analyze the dynamics of the $S=1$ Kitaev spin liquid within this time-reversal-symmetric ansatz and discuss its physical properties.
We also compute the temperature evolution of the dynamical spin structure factor using the latter scheme.
We find that the continuum observed at zero temperature splits into two distinct structures as the temperature increases.
We clarify that this splitting behavior arises from modifications in the spinon dispersion caused by thermal fluctuations of the mean fields.
This framework can also be applied to other quantum spin models to provide deeper insights into the spin dynamics of QSLs at finite temperatures.

This paper is organized as follows. In Sec.~\ref{sec:Method}, we present the method employed in this study.
The Schwinger boson theory and the mean-field approximation applied to it are described in Secs.~\ref{sec:Schwinger boson theory} and \ref{sec:Mean-field theory}, respectively.
Section~\ref{sec:Calculation of the spin structure factor} outlines the calculation schemes for the dynamical spin correlations based on SBMFT.
We propose two distinct schemes according to the decoupling procedures: one based on the conventional decoupling with respect to the spinon operators, and the other formulated in terms of the bond operators.
In Sec.~\ref{sec:Model}, we present the $S=1$ Kitaev model on a honeycomb lattice.
Section~\ref{sec:results} presents the results obtained in this work.
In Sec.~\ref{sec:Mean-field ansatz}, we provide the mean-field ansatz employed in our calculations.
In Sec.~\ref{sec:Zero-temperature}, we present the zero-temperature results: the dynamical spin structure factor $S(\bm{q},\omega)$ and the equal-time structure factor $S(\bm{q})$ for both the antiferromagnetic and ferromagnetic Kitaev models.
We also demonstrate that the two evaluation schemes introduced in Sec.~\ref{sec:Calculation of the spin structure factor} yield qualitatively different behaviors.
In Sec.~\ref{sec:Finite-temperature}, we show the results for the finite-temperature evolution of the dynamical spin structure factor $S(\bm{q},\omega)$.
In Sec.~\ref{sec:Discussion}, we discuss the origin of the discrepancies between the two decoupling schemes.
Finally, we summarize our findings in Sec.~\ref{sec:Summary}.

\section{Method}
\label{sec:Method}

\subsection{Schwinger boson theory}
\label{sec:Schwinger boson theory}

In this section, we provide an overview of the Schwinger boson theory for quantum spin systems~\cite{Arovas-Auerbach-1998,Read-Sachdev-1991,Sachdev-Read-1991,Sachdev-1992} and its extension to generic interactions.
We first introduce the conventional framework based on SU($2$)-invariant bond operators and then describe our formulation, which extends this framework to include SU($2$)-breaking operators, thereby enabling a unified description of arbitrary spin-spin interactions.

In the Schwinger boson representation, the $\gamma\left(=x,y,z\right)$ component of spin operators with the length $S$ at each site $i$ defined on a lattice is expressed in terms of a pair of bosons, $\bm{b}_{i}=\left(b_{i\uparrow},b_{i\downarrow}\right)^T$, as
\begin{align}
    \label{eq:Schwinger boson representation}
    S_{i}^{\gamma}
    =\frac{1}{2}\sum_{\mu,\nu=\uparrow,\downarrow}b_{i\mu}^{\dagger}\sigma_{\mu\nu}^{\gamma}b_{i\nu},
\end{align}
where $\bm{\sigma}=(\sigma^{x},\sigma^{y},\sigma^{z})$ denotes a set of Pauli matrices.
The operators $b_{i\uparrow}$ and $b_{i\downarrow}$ satisfy the following bosonic commutation relation:
\begin{align}
    \label{eq:commutation relation}
    \left[
    b_{i\mu},b_{j\nu}^{\dagger}
    \right]=\delta_{ij}\delta_{\mu\nu}.
\end{align}
Hereafter, we refer to the bosonic quasiparticles described by $b_{i\uparrow}$ and $b_{i\downarrow}$ as spinons.
Because this bosonic representation enlarges the Hilbert space, it is necessary to impose the local constraint,
\begin{align}
    \label{eq:local constraint}
    n_{i}=\sum_{\mu}b_{i\mu}^{\dagger}b_{i\mu}=2S,
\end{align}
which projects the Hilbert space described by the bosonic operators, $b_{i\uparrow}$ and $b_{i\downarrow}$, onto the physical subspace and ensures that the spin length satisfies $\bm{S}^{2}=S(S+1)$.
Only if the constraint is enforced exactly on every lattice site, the Schwinger boson formulation yields exact results.

Thus far, the Schwinger boson representation has primarily been applied to spin systems involving the Heisenberg interaction.
To describe this interaction, the following two bond operators, $\mathcal{B}_{ij}$ and $\mathcal{A}_{ij}$, are introduced as
\begin{align}
    \label{eq:SU(2) invariant operators B}
    \mathcal{B}_{ij}
    =
    \frac{1}{2}\sum_{\mu,\nu}b_{i\mu}^{\dagger}\sigma^{0}_{\mu\nu}b_{j\nu}
    =\frac{1}{2}\left(
    b_{i\uparrow}^{\dagger}b_{j\uparrow}+b_{i\downarrow}^{\dagger}b_{j\downarrow}
    \right),
\end{align}
and
\begin{align}
    \label{eq:SU(2) invariant operators A}
    \mathcal{A}_{ij}
    &=
    \frac{i}{2}\sum_{\mu,\nu}b_{i\mu}\sigma^{y}_{\mu\nu}b_{j\nu}=\frac{1}{2}\left(b_{i\uparrow}b_{j\downarrow}-b_{i\downarrow}b_{j\uparrow}\right),
\end{align}
respectively, where $\sigma^{0}_{\mu\nu}$ denotes the $2\times 2$ identity matrix.
Note here that both $\mathcal{B}_{ij}$ and $\mathcal{A}_{ij}$ remain invariant under global SU($2$) rotations of the two-dimensional vector $\bm{b}_{i}=\left(b_{i\uparrow},b_{i\downarrow}\right)^T$.
Physically, $\mathcal{B}_{ij}$ represents spinon hopping, whereas $\mathcal{A}_{ij}$ describes the resonating spin-singlet amplitude on the bond $\braket{i,j}$.
Using the identity $\sum_{\gamma=x,y,z}\sigma_{\mu\nu}^{\gamma}\sigma_{\rho\lambda}^{\gamma}=2\sigma_{\mu\lambda}^{0}\sigma_{\nu\rho}^{0}-\sigma_{\mu\nu}^{0}\sigma_{\rho\lambda}^{0}$,
the Heisenberg interaction $\bm{S}_{i}\cdot\bm{S}_{j}$ can be expressed in terms of the bond operator $\mathcal{B}_{ij}$ as
\begin{align}
    \label{eq:Heinsenberg interaction Schwinger boson representation-B}
    \bm{S}_{i}\cdot\bm{S}_{j}
    &=2:\mathcal{B}_{ij}^{\dagger}\mathcal{B}_{ij}:-S^{2},
\end{align}
where $:\mathcal{O}_{1}\mathcal{O}_{2}:$ denotes normal ordering, in which all creation operators are arranged to the left of annihilation operators.
Since the relation $\sum_{\gamma=x,y,z}\sigma_{\mu\nu}^{\gamma}\sigma_{\rho\lambda}^{\gamma}=\sigma_{\mu\nu}^{0}\sigma_{\rho\lambda}^{0}+2\sigma_{\mu\rho}^{y}\sigma_{\nu\lambda}^{y}$ also holds, the Heisenberg interaction can alternatively be rewritten as
\begin{align}
    \label{eq:Heinsenberg interaction Schwinger boson representation-A}
    \bm{S}_{i}\cdot\bm{S}_{j}
    =S^{2}-2\mathcal{A}_{ij}^{\dagger}\mathcal{A}_{ij}.
\end{align}

Originally, the representation of the Heisenberg interaction using Eq.~\eqref{eq:Heinsenberg interaction Schwinger boson representation-B} with the $\mathcal{B}_{ij}$ operators had been applied only to the ferromagnetic case, while the representation given in Eq.~\eqref{eq:Heinsenberg interaction Schwinger boson representation-A} with the $\mathcal{A}_{ij}$ operators had been used for the antiferromagnetic case~\cite{Arovas-Auerbach-1998}. However, subsequent studies~\cite{
Gazza-1993,
Manuel-Trumper-1994,
Mattsson-1995,
Flint-Coleman-2009,Isakov-2009,
Mezio-2011,
Messio-Bernu-2012,
Messio-Lhuillier-2013,
Merino-Holt-Powell-2014,
Halimeh-Punk-2016,
Bauer-2017,Messio-Bieri-2017,Mondal-2017,Halimeh-Singh-2019,
Gonzalez-2017,Chern-Hwang-2017,Messio-Bieri-2017,Mondal-2017,
Ghioldi-Gonzalez-2018,
Zhang-Ghioldi-2019,Halimeh-Singh-2019,Mondal-2019,
Gonzalez-2020,
Zhang-2021,Mondal-2021,
Ghioldi-Zhang-2022,Li-Li-2022,Lugan-Jaubert-2022,
Rossi-Motruk-2023,
Pomponio-2024} have revealed that, in various frustrated magnets, the properties of the ground state and the excitation spectra can be well captured by employing the symmetrized form given as follows:
\begin{align}
    \label{eq:Heinsenberg interaction Schwinger boson representation}
    \bm{S}_{i}\cdot\bm{S}_{j}
    =:\mathcal{B}_{ij}^{\dagger}\mathcal{B}_{ij}:-\mathcal{A}_{ij}^{\dagger}\mathcal{A}_{ij},
\end{align}
which is obtained by averaging the two representations in Eqs.~\eqref{eq:Heinsenberg interaction Schwinger boson representation-B} and \eqref{eq:Heinsenberg interaction Schwinger boson representation-A}.
In light of these findings, we adopt this mixed representation throughout the present work.

To extend the Schwinger boson representation to include generic spin interactions, we introduce a set of bond operators that break the SU($2$) symmetry~\cite{
Kargarian-Langari-Fiete-2012,
Kos-Punk-2017,
Samajdar-Scheurer-2019,
Mondal-2021,
Schneider-2022,
Ralko-Merino-2024}, in addition to the SU($2$)-invariant bond operators $\mathcal{B}_{ij}$ and $\mathcal{A}_{ij}$.
As additional SU($2$)-breaking bond operators, we define the following operators:
\begin{align}
    \label{eq:SU(2) breaking operators C}
    \mathcal{C}_{ij}^{\gamma}
    &=
    \frac{1}{2}\sum_{\mu,\nu}b_{i\mu}^{\dagger}\sigma^{\gamma}_{\mu\nu}b_{j\nu},
    \\
    \label{eq:SU(2) breaking operators D}
    \mathcal{D}_{ij}^{\gamma}
    &=
    \frac{i}{2}\sum_{\mu,\nu}b_{i\mu}\left(\sigma^{y}\sigma^{\gamma}\right)_{\mu\nu}b_{j\nu}.
\end{align}
By employing these eight bond operators, $\mathcal{A}_{ij}$, $\mathcal{B}_{ij}$, $\mathcal{C}_{ij}^{x}$, $\mathcal{C}_{ij}^{y}$, $\mathcal{C}_{ij}^{z}$, $\mathcal{D}_{ij}^{x}$, $\mathcal{D}_{ij}^{y}$, and $\mathcal{D}_{ij}^{z}$, any intersite bilinear bosonic term can be expressed, and hence any bilinear spin interaction can be represented using these bond operators.
Here, the spin interaction between two sites $i$ and $j$ involving different spin components is represented as
\begin{align}
    \label{eq:general spin interaction Schwinger boson representation}
    S_{i}^{\alpha}S_{j}^{\beta}=\sum_{p,q}\left(A_{pq}^{\alpha\beta}:\mathcal{Q}_{ij}^{p\dagger}\mathcal{Q}_{ij}^{q}:\right),
\end{align}
where $\bm{\mathcal{Q}}_{ij}=
\left(
    \mathcal{A}_{ij}, \mathcal{B}_{ij}, \mathcal{C}_{ij}^{x}, \mathcal{C}_{ij}^{y}, \mathcal{C}_{ij}^{z}, \mathcal{D}_{ij}^{x}, \mathcal{D}_{ij}^{y}, \mathcal{D}_{ij}^{z}
\right)$ comprises both the SU($2$)-invariant and SU($2$)-breaking bond operators, and $A_{pq}^{\alpha\beta}$ are coefficients that depend on the spin components $\alpha$ and $\beta$.
The indices $p, q=1,\cdots,8$ label the components of $\bm{\mathcal{Q}}_{ij}$.
The coefficients $A_{pq}^{\alpha\beta}$ are not unique, because the same spin interaction can be represented by different, algebraically equivalent bond-operator identities before the mean-field approximation is applied.
In the following, we choose the symmetrized bond-operator representation, in analogy with the symmetrized Heisenberg form in Eq.~\eqref{eq:Heinsenberg interaction Schwinger boson representation}.
Note that the following relations hold under the exchange of site indices: $\mathcal{A}_{ji}=-\mathcal{A}_{ij}$, $\mathcal{B}_{ji}=\mathcal{B}_{ij}^\dagger$, $\mathcal{C}_{ji}^{\gamma}=\mathcal{C}_{ij}^{\gamma\dagger}$, and $\mathcal{D}_{ji}^{\gamma}=\mathcal{D}_{ij}^{\gamma}$.

In the Kitaev model on a honeycomb lattice introduced later, the interaction is bond-dependent and of the Ising type, $S_{i}^{\gamma}S_{j}^{\gamma}$, which can be expressed in the present formalism as
\begin{align}
    \label{eq:Ising interaction Schwinger boson representation}
    S_{i}^{\gamma}S_{j}^{\gamma}=\frac{1}{2}\left(:\mathcal{B}_{ij}^{\dagger}\mathcal{B}_{ij}:-\mathcal{D}_{ij}^{\gamma\dagger}\mathcal{D}_{ij}^{\gamma}+:\mathcal{C}_{ij}^{\gamma\dagger}\mathcal{C}_{ij}^{\gamma}:-\mathcal{A}_{ij}^{\dagger}\mathcal{A}_{ij}\right).
\end{align}
For example, the $S_{i}^{x}S_{j}^{x}$ interaction is represented by choosing the coefficients in Eq.~\eqref{eq:general spin interaction Schwinger boson representation} as $A_{11}^{xx}=-1/2,~A_{22}^{xx}=+1/2,~A_{33}^{xx}=+1/2,~A_{66}^{xx}=-1/2$, with all other $A_{pq}^{xx}$ set to zero.
The $S_{i}^{y}S_{j}^{y}$ and $S_{i}^{z}S_{j}^{z}$ interactions can be expressed in a similar manner by assigning the appropriate coefficients in Eq.~\eqref{eq:general spin interaction Schwinger boson representation}.
A detailed derivation of Eq.~\eqref{eq:Ising interaction Schwinger boson representation} is provided in Appendix A, where we also present the Schwinger boson representations of the off-diagonal spin interactions $S_{i}^{\mu}S_{j}^{\nu}$ with $\mu\neq\nu$.

\subsection{Mean-field theory}
\label{sec:Mean-field theory}

We begin with a generic quantum spin Hamiltonian expressed by
\begin{align}
  \mathcal{H}
  =\frac{1}{2}\sum_{i,j}J_{ij}^{\alpha\beta}S_{i}^{\alpha}S_{j}^{\beta},
\end{align}
where $\bm{S}_i$ is defined at each site on a lattice with $N$ spins. $J_{ij}^{\alpha\beta}$ denotes the coefficient of exchange interactions between spins at sites $i$ and $j$ in the $\alpha$ and $\beta$ components, respectively, and satisfies the relation $J_{ij}^{\alpha\beta}=J_{ji}^{\beta\alpha}$.
In the previous section, we introduced the generalized Schwinger boson representation of the spin operators, which enables us to describe any spin interactions as the product of two bond operators, as
\begin{align}
  \label{eq:spin Hamiltonian Schwinger representation}
  \mathcal{H}
  =\frac{1}{2}\sum_{i,j}\sum_{p,q}J_{ij}^{\alpha\beta}\left(A_{pq}^{\alpha\beta}:\mathcal{Q}_{ij}^{p\dagger}\mathcal{Q}_{ij}^{q}:\right).
\end{align}
Although this representation is exact as long as the local constraint in Eq.~\eqref{eq:local constraint} is imposed, each term includes four boson operators corresponding to interactions between bosons. Therefore, approximations are required within the Schwinger boson theory.
To address these interactions, we apply the mean-field approximation to Eq.~\eqref{eq:spin Hamiltonian Schwinger representation}, which is known as SBMFT.
In this approach, each bond operator is decomposed into its expectation value and the fluctuation around it, as
\begin{align}
    \label{eq:deviation}
    \mathcal{Q}_{ij}^{\gamma}=\delta \mathcal{Q}_{ij}^{\gamma} + \braket{\mathcal{Q}_{ij}^{\gamma}},
\end{align}
where we assume that the thermal average $\braket{\mathcal{Q}_{ij}^{\gamma}}$ possesses a periodicity commensurate with the lattice structure, and introduce the superlattice unit cell reflecting this periodicity.
In the present formulation, we consider that $M$ sites are included in the superlattice unit cell.

By using Eq.~\eqref{eq:deviation}, the Hamiltonian in Eq.~\eqref{eq:spin Hamiltonian Schwinger representation} can be decomposed into the mean-field Hamiltonian $\mathcal{H}^{\text{MF}}_{ij}$ and the deviation term $\mathcal{H}^{\prime}$ as
\begin{align}
    \label{eq:mean field Hamiltonian}
    \mathcal{H}=\sum_{i,j}\mathcal{H}_{ij}^{\text{MF}}+\mathcal{H}^{\prime},
\end{align}

where the mean-field Hamiltonian $\mathcal{H}_{ij}^{\text{MF}}$ on the bond connecting sites $i$ and $j$ is given by
\begin{align}
    \label{eq:local Hamiltonian}
    &\mathcal{H}_{ij}^{\text{MF}}=\notag\\
    &\frac{J^{\alpha\beta}_{ij}}{2}\left(\sum_{p,q}A_{pq}^{\alpha\beta}\left[\braket{\mathcal{Q}_{ij}^{p\dagger}}\mathcal{Q}_{ij}^{q}+\braket{\mathcal{Q}_{ij}^{q}}\mathcal{Q}_{ij}^{p\dagger}-\braket{\mathcal{Q}_{ij}^{p\dagger}}\braket{\mathcal{Q}_{ij}^{q}}\right]\right).
\end{align}
In the mean-field treatment below, we neglect the deviation term $\mathcal{H}^{\prime}$.
Furthermore, it is necessary to impose the local constraint in Eq.~\eqref{eq:local constraint} in the self-consistent calculations.
However, it is practically challenging to enforce the local constraint at each site. 
Instead, we introduce a uniform Lagrange multiplier term as
$ \lambda\sum_{i}\left(n_{i}-2S\right).$
This term enforces a constraint on the average number of bosons per site by incorporating it into the Hamiltonian.
The resulting mean-field Hamiltonian is expressed as
\begin{align}
    \label{eq:MF Hamiltonian}
    \mathcal{H}^{\text{MF}}=\sum_{i,j}\mathcal{H}_{ij}^{\text{MF}}+\lambda\sum_{i}\left(n_{i}-2S\right).
\end{align}
Since the bond operator $\bm{\mathcal{Q}}_{ij}$ can be expressed in terms of the bosonic operators $\bm{b}_{i}$ and $\bm{b}_{j}$, the mean-field Hamiltonian $\mathcal{H}^{\text{MF}}$ can be written in the bilinear form as
\begin{align}
    \label{eq:BdG Hamiltonian in real space}
    \mathcal{H}^{\text{MF}}=\frac{1}{2}\sum_{i,j}B_{l}^{\dagger}\mathcal{M}_{ll'}B_{l'}+\text{const},
\end{align}
where site $i$ is labeled by the unit cell index $l=1,2,\dots,N/M$ and the sublattice index $m=1,2,\dots, M$ as $i=(l,m)$, $\mathcal{M}_{ll'}$ is a $4M\times 4M$ Hermitian matrix, and $B_{l}^{\dagger}$ is a $4M$-dimensional vector defined as
\begin{align}
    \label{eq:def B_{i}}
    B_{l}^{\dagger}
    =
    \left(
    b_{l,1,\uparrow}^{\dagger}\cdots b_{l,M,\uparrow}^{\dagger},b_{l,1,\downarrow}^{\dagger}\cdots b_{l,M,\downarrow}^{\dagger},b_{l,1,\uparrow}\cdots b_{l,M,\uparrow},b_{l,1,\downarrow}\cdots b_{l,M,\downarrow}
    \right).
\end{align}
Hereafter, we drop additive constants in Eq.~\eqref{eq:BdG Hamiltonian in real space}, which only shift the reference energy and do not affect the Heisenberg time evolution or spin correlation functions.

By applying the Fourier transformation to the bosonic operators, we obtain the momentum-space representation of the mean-field Hamiltonian as
\begin{align}
    \label{eq:BdG Hamiltonian in momentum space}
    \mathcal{H}^{\text{MF}}=\frac{1}{2}\sum_{\bm{k}}^{\text{B.Z.}}B_{\bm{k}}^{\dagger}\mathcal{M}_{\bm{k}}B_{\bm{k}},
\end{align}
where the sum over $\bm{k}$ is taken in the first Brillouin zone of the superlattice, and $B_{\bm{k}}^{\dagger}$ is the Fourier-transformed vector defined as
\begin{align}
    \label{eq:def B_{k}}
    B_{\bm{k}}^{\dagger}
    =
    \Bigl(&
    b_{\bm{k},1,\uparrow}^{\dagger}\cdots b_{\bm{k},M,\uparrow}^{\dagger},b_{\bm{k},1,\downarrow}^{\dagger}\cdots b_{\bm{k},M,\downarrow}^{\dagger},\notag\\
    &\quad b_{-\bm{k},1,\uparrow}\cdots b_{-\bm{k},M,\uparrow},b_{-\bm{k},1,\downarrow}\cdots b_{-\bm{k},M,\downarrow}
    \Bigr).
\end{align}
with
\begin{align}
    \label{eq:Fourier transformation}
    b_{\bm{k},m}^\dagger
    &=\sqrt{\frac{M}{N}}\sum_{l} b_{(l,m)}^\dagger e^{i\bm{k}\cdot\bm{R}_{l}}.
\end{align}
Here, $\bm{R}_{l}$ is the representative position of unit cell $l$, and  $\mathcal{M}_{\bm{k}}$ is a $4M\times 4M$ matrix defined by the Fourier transformation of the real-space matrix $\mathcal{M}_{ll'}$ as
\begin{align}
    \label{eq:def M_{k}}
    \mathcal{M}_{\bm{k}}=\sum_{l'}\mathcal{M}_{ll'}e^{-i\bm{k}\cdot\left(\bm{R}_{l}-\bm{R}_{l'}\right)}.
\end{align}
Note that $\mathcal{M}_{\bm{k}}$ is independent of $l$ due to translational symmetry, implying that $\mathcal{M}_{ll'}$ depends only on the relative position $\bm{R}_{l}-\bm{R}_{l'}$.

We diagonalize $\mathcal{M}_{\bm{k}}$ via the Bogoliubov transformation as $\mathcal{E}_{\bm{k}}=\mathcal{T}_{\bm{k}}^{\dagger}\mathcal{M}_{\bm{k}}\mathcal{T}_{\bm{k}}$, where $\mathcal{T}_{\bm{k}}$ is a $4M\times 4M$ paraunitary matrix that satisfies the relation $\mathcal{T}_{\bm{k}}^{\dagger}\sigma_{3}\mathcal{T}_{\bm{k}}=\mathcal{T}_{\bm{k}}\sigma_{3}\mathcal{T}_{\bm{k}}^{\dagger}=\sigma_{3}$.
Here, we introduce the paraunit matrix
$\left(\begin{smallmatrix}\bm{1}_{2M\times 2M}&0\\0& -\bm{1}_{2M\times 2M}
  \end{smallmatrix}\right)$, where $\bm{1}_{2M\times 2M}$ denotes the $2M\times 2M$ identity matrix, and $\mathcal{E}_{\bm{k}}$ is the diagonal matrix given by $\mathcal{E}_{\bm{k}}=\text{diag}\left\{
\varepsilon_{\bm{k},1},\cdots,\varepsilon_{\bm{k},2M},\varepsilon_{-\bm{k},1},\cdots,\varepsilon_{-\bm{k},2M}
\right\}$.
Using this transformation, the mean-field Hamiltonian can be expressed in the following diagonalized form:
\begin{align}
    \label{eq:diagonalized form}
    \mathcal{H}^{\text{MF}}
    =\frac{1}{2}\sum_{\bm{k}}^{\text{B.Z.}}\Gamma_{\bm{k}}^{\dagger}\mathcal{E}_{\bm{k}}\Gamma_{\bm{k}},
\end{align}
Here, we introduce the set of bosonic operators $\Gamma_{\bm{k}}=\mathcal{T}_{\bm{k}}^{-1}B_{\bm{k}}$,
which is defined by
\begin{align}
    \label{eq:def Gamma_{k}}
    \Gamma_{\bm{k}}
    =\left(
    \gamma_{\bm{k},1}^{\dagger}\cdots\gamma_{\bm{k},2M}^{\dagger},\gamma_{-\bm{k},1}\cdots\gamma_{-\bm{k},2M}
    \right),
\end{align}
where $\gamma_{\bm{k},\eta}$ and $\gamma_{\bm{k},\eta}^{\dagger}$ are the annihilation and creation operators of a bosonic quasiparticle with energy $\varepsilon_{\bm{k},\eta}$.
We can evaluate the expectation values of the bond operators $\mathcal{Q}_{ij}^{\gamma}$ and the number operator $n_i$ in Eq.~\eqref{eq:local constraint} using $\braket{\gamma_{\bm{k},\eta}^\dagger\gamma_{\bm{k},\eta}}=g(\varepsilon_{\bm{k},\eta})$, where $g(\varepsilon)=1/(e^{\varepsilon/T}-1)$ denotes the Bose distribution function at temperature $T$, assuming that the Boltzmann constant $k_{\rm B}$ is set to unity.
Since the mean-field Hamiltonian $\mathcal{H}^{\text{MF}}$ depends on the mean-field parameters $\braket{\mathcal{Q}_{ij}^{\gamma}}$ and the Lagrange multiplier $\lambda$, the values of $\braket{\mathcal{Q}_{ij}^{\gamma}}$ are determined self-consistently, and the Lagrange multiplier $\lambda$ is determined such that $\sum_i\braket{n_i}=2NS$ holds.
Note that both the Lagrange multiplier $\lambda$ and the mean-field parameters $\langle \mathcal{Q}^{\gamma}_{ij}\rangle$ are temperature dependent, and are determined self-consistently at each temperature.

\subsection{Calculation of the spin structure factor}
\label{sec:Calculation of the spin structure factor}

Within the SBMFT, we calculate the dynamical spin structure factor.
Throughout this subsection, expectation values and time evolution are evaluated with respect to the self-consistent quadratic mean-field Hamiltonian $\mathcal{H}^{\text{MF}}$.
The dynamical spin structure factor is defined as the Fourier transform of the space-time spin correlators and is given by
\begin{align}
    \label{eq:dynamical spin structure factor}
    S^{\alpha\alpha^{\prime}}(\bm{q},\omega) = \frac{1}{N}\int_{-\infty}^{\infty} dt e^{i\omega t}\sum_{i,j} e^{-i\bm{q}\cdot\left(\bm{r}_{i}-\bm{r}_{j}\right)} \braket{S^{\alpha}_{i}(t) S^{\alpha^{\prime}}_{j}},
\end{align}
Here, the real-time spin operator is evaluated in the Heisenberg picture with the mean-field Hamiltonian as
\begin{align}
    \label{eq:spin-operator-time-dependence}
    S^{\alpha}_{i}(t)
    =
    \frac{1}{2}
    \sum_{\mu,\nu}
    b_{i\mu}^{\dagger}(t)
    \sigma_{\mu\nu}^{\alpha}
    b_{i\nu}(t),
\end{align}
where
\begin{align}
    b_{i\mu}^{\dagger}(t)
    &=
    e^{i\mathcal{H}^{\text{MF}}t}b_{i\mu}^{\dagger}e^{-i\mathcal{H}^{\text{MF}}t},
    \label{eq:create-boson-operator-time-dependence}
    \\
    b_{i\nu}(t)
    &=
    e^{i\mathcal{H}^{\text{MF}}t}b_{i\nu}e^{-i\mathcal{H}^{\text{MF}}t}.
    \label{eq:annihilate-boson-operator-time-dependence}
\end{align}
The equal-time spin structure factor is obtained by integrating the dynamical one over frequency:
\begin{align}
    \label{eq:static spin structure factor}
    S^{\alpha\alpha^{\prime}}(\bm{q}) = \int_{-\infty}^{\infty} \frac{d\omega}{2\pi} S^{\alpha\alpha^{\prime}}(\bm{q},\omega) = \frac{1}{N}\sum_{i,j} \braket{S_{i}^{\alpha} S_{j}^{\alpha^{\prime}}} e^{-i\bm{q}\cdot(\bm{r}_{i}-\bm{r}_{j})},
\end{align}
which coincides with the Fourier transform of the equal-time correlation function $\braket{S_{i}^{\alpha} S_{j}^{\alpha^{\prime}}}$.
To evaluate the spin structure factor in the SBMFT, we first formulate the corresponding correlation function in imaginary time and then analytically continue the result to real frequencies.
We define
\begin{align}
    \label{eq:imaginary-time-spin-correlation}
    \chi_{ij}^{\alpha\alpha^{\prime}}(\tau)
    =
    \braket{T_{\tau}S_{i}^{\alpha}(\tau)S_{j}^{\alpha^{\prime}}(0)},
    \qquad
    0\leq \tau<\beta,
\end{align}
where $T_{\tau}$ denotes imaginary-time ordering and $\beta=1/T$.
The imaginary-time spin operator is written as
\begin{align}
    \label{eq:imaginary-time-spin-operator}
    S_{i}^{\alpha}(\tau)
    =
    \frac{1}{2}
    \sum_{\mu,\nu}
    \bar{b}_{i\mu}(\tau)\sigma_{\mu\nu}^{\alpha}b_{i\nu}(\tau),
\end{align}
where
\begin{align}
    \label{eq:imaginary-time-boson-operators}
    \bar{b}_{i\mu}(\tau)
    &=
    e^{\tau\mathcal{H}^{\text{MF}}}b_{i\mu}^{\dagger}e^{-\tau\mathcal{H}^{\text{MF}}},
    &
    b_{i\mu}(\tau)
    &=
    e^{\tau\mathcal{H}^{\text{MF}}}b_{i\mu}e^{-\tau\mathcal{H}^{\text{MF}}}.
\end{align}
Here, the bar on $\bar{b}_{i\mu}(\tau)$ is only a notation for the imaginary-time-evolved creation operator and should not be interpreted as the Hermitian conjugate of $b_{i\mu}(\tau)$.
The Matsubara susceptibility is
\begin{align}
    \label{eq:matsubara-spin-susceptibility}
    \chi^{\alpha\alpha^{\prime}}(\bm{q},i\omega_{n})
    =
    \frac{1}{N}
    \sum_{i,j}
    e^{-i\bm{q}\cdot(\bm{r}_{i}-\bm{r}_{j})}
    \int_{0}^{\beta}d\tau\,
    e^{i\omega_{n}\tau}
    \chi_{ij}^{\alpha\alpha^{\prime}}(\tau),
\end{align}
where $\omega_{n}=2\pi n/\beta$ is a bosonic Matsubara frequency.
The dynamical structure factor is obtained by analytically continuing
$i\omega_n$ to $\omega+i\delta$ as
\begin{align}
    \label{eq:fluctuation-dissipation}
    S^{\alpha\alpha^{\prime}}(\bm{q},\omega)
    =
    \frac{2}{1-e^{-\beta\omega}}
    \mathrm{Im}\,
    \chi^{\alpha\alpha^{\prime}}(\bm{q},i\omega_{n}\rightarrow\omega+i\delta),
\end{align}
where $\delta>0$ is taken to be infinitesimal analytically and is replaced by a small broadening in numerical calculations.
The equal-time structure factor can equivalently be obtained from
\begin{align}
    \label{eq:static-spin-structure-factor-matsubara}
    S^{\alpha\alpha^{\prime}}(\bm{q})
    =
    \frac{1}{\beta}
    \sum_{n}
    \chi^{\alpha\alpha^{\prime}}(\bm{q},i\omega_{n}).
\end{align}

Since $\mathcal{H}^{\text{MF}}$ is quadratic, all two-point contractions appearing below are evaluated by diagonalizing $\mathcal{H}^{\text{MF}}$ and carrying out the Matsubara sums.

We now specify the two decoupling schemes used to evaluate $\chi_{ij}^{\alpha\alpha^{\prime}}(\tau)$.
The conventional scheme, referred to as decoupling I, performs the Wick decomposition directly in terms of the Schwinger bosons.
Since we focus on a quantum spin-liquid state with no long-range magnetic order, the local spin expectation values vanish, $\braket{S_{i}^{\alpha}}=\braket{S_{j}^{\alpha^{\prime}}}=0$, and the disconnected one-point contribution is absent.
The imaginary-time version of decoupling I is
\begin{align}
      \label{eq:Wick decoupling}
      \chi_{ij,\mathrm{I}}^{\alpha\alpha^{\prime}}(\tau)
      &=
      \frac{1}{4}
      \sum_{\mu,\nu,\rho,\lambda}
      \sigma_{\mu\nu}^{\alpha}
      \sigma_{\rho\lambda}^{\alpha^{\prime}}
      \nonumber\\
      &\quad\times
      \Bigl[
      \braket{T_{\tau}\bar{b}_{i\mu}(\tau)b_{j\lambda}(0)}
      \braket{T_{\tau}b_{i\nu}(\tau)\bar{b}_{j\rho}(0)}
      \nonumber\\
      &\qquad\qquad
      +
      \braket{T_{\tau}\bar{b}_{i\mu}(\tau)\bar{b}_{j\rho}(0)}
      \braket{T_{\tau}b_{i\nu}(\tau)b_{j\lambda}(0)}
      \Bigr].
\end{align}

This decoupling I was introduced in the original paper that proposed the SBMFT for quantum magnets to calculate spin correlations~\cite{Arovas-Auerbach-1998}.
Consequently, many subsequent studies employing SBMFT have adopted the same scheme~\cite{
Lefmann-Hedegard-1994,
Mezio-2011,
Merino-Holt-Powell-2014,
Halimeh-Punk-2016,
Bauer-2017,Mondal-2017,
Kos-Punk-2017,
Halimeh-Singh-2019,
Mondal-2021,
Schneider-2022,
Rossi-Motruk-2023,
Ralko-Merino-2024}.
This scheme retains all normal and anomalous spinon contractions generated by the quadratic mean-field Hamiltonian.
However, we demonstrate here that this conventional choice can yield a momentum dependence qualitatively different from that expected from the exchange interaction.

The second scheme, decoupling II, is instead formulated in the bond-operator language used in the mean-field decoupling of the Hamiltonian.
The starting point is the bond-operator identity
\begin{align}
    \label{eq:bond-operator-identity-for-correlations}
    S_{i}^{\alpha}S_{j}^{\alpha^{\prime}}
    =
    \sum_{p,q}
    A_{pq}^{\alpha\alpha^{\prime}}
    :
    \mathcal{Q}_{ij}^{p\dagger}
    \mathcal{Q}_{ij}^{q}
    :,
\end{align}
where $p,q=1,\cdots,8$ label the components of $\bm{\mathcal{Q}}_{ij}$.
At unequal imaginary times, we denote the corresponding time-split vector by $\mathscr{Q}^{p}_{ij}(\tau)$; its components are $\mathscr{A}_{ij}(\tau)$, $\mathscr{B}_{ij}(\tau)$, $\mathscr{C}_{ij}^{\gamma}(\tau)$, and $\mathscr{D}_{ij}^{\gamma}(\tau)$ with $\gamma=x,y,z$, ordered in the same way as the corresponding components of $\bm{\mathcal{Q}}_{ij}$.
Each component is defined by placing the operator on site $i$ at time $\tau$ and that on site $j$ at time $0$:
\begin{subequations}
\label{eq:time-split-bond-contractions}
\begin{align}
    \mathscr{A}_{ij}(\tau)
    &=
    \frac{i}{2}
    \sum_{\mu,\nu}
    \sigma_{\mu\nu}^{y}
    \braket{T_{\tau}b_{i\mu}(\tau)b_{j\nu}(0)},
    \label{eq:time-split-bond-contraction-A}
    \\
    \mathscr{B}_{ij}(\tau)
    &=
    \frac{1}{2}
    \sum_{\mu}
    \braket{T_{\tau}\bar{b}_{i\mu}(\tau)b_{j\mu}(0)},
    \label{eq:time-split-bond-contraction-B}
    \\
    \mathscr{C}_{ij}^{\gamma}(\tau)
    &=
    \frac{1}{2}
    \sum_{\mu,\nu}
    \sigma_{\mu\nu}^{\gamma}
    \braket{T_{\tau}\bar{b}_{i\mu}(\tau)b_{j\nu}(0)},
    \label{eq:time-split-bond-contraction-C}
    \\
    \mathscr{D}_{ij}^{\gamma}(\tau)
    &=
    \frac{i}{2}
    \sum_{\mu,\nu}
    (\sigma^{y}\sigma^{\gamma})_{\mu\nu}
    \braket{T_{\tau}b_{i\mu}(\tau)b_{j\nu}(0)}.
    \label{eq:time-split-bond-contraction-D}
\end{align}
\end{subequations}
The barred contraction $\bar{\mathscr{Q}}_{ij}^{p}(\tau)$ is defined analogously by applying the same time-split prescription to the Hermitian-conjugate bond operator $\mathcal{Q}_{ij}^{p\dagger}$.
With this definition, decoupling II is obtained by factorizing the bond-operator product in Eq.~\eqref{eq:bond-operator-identity-for-correlations} into the time-split bond contractions:
\begin{align}
  \label{eq:mean-field decoupling}
  \chi_{ij,\mathrm{II}}^{\alpha\alpha^{\prime}}(\tau)
  &=
  \sum_{p,q}
  A_{pq}^{\alpha\alpha^{\prime}}
  \bar{\mathscr{Q}}_{ij}^{p}(\tau)
  \mathscr{Q}_{ij}^{q}(\tau).
\end{align}
For the diagonal Ising component relevant to a $\gamma$ Kitaev bond, this becomes
\begin{align}
    \label{eq:decoupling-II-kitaev-component}
    \chi_{ij,\mathrm{II}}^{\gamma\gamma}(\tau)
    &=
    \frac{1}{2}
    \Bigl[
    \bar{\mathscr{B}}_{ij}(\tau)\mathscr{B}_{ij}(\tau)
    -
    \bar{\mathscr{D}}_{ij}^{\gamma}(\tau)\mathscr{D}_{ij}^{\gamma}(\tau)
    \nonumber\\
    &\qquad\qquad
    +
    \bar{\mathscr{C}}_{ij}^{\gamma}(\tau)\mathscr{C}_{ij}^{\gamma}(\tau)
    -
    \bar{\mathscr{A}}_{ij}(\tau)\mathscr{A}_{ij}(\tau)
    \Bigr].
\end{align}
Thus, decoupling II does not apply Wick decomposition directly to the Schwinger-boson four-point function.
Instead, it first rewrites the spin product in terms of bond operators and then factorizes the correlator in the same bond channels that define the SBMFT saddle point.

For either decoupling scheme, the Matsubara sums are carried out after expressing the time-split contractions in terms of the eigenmodes of the quadratic mean-field Hamiltonian.
We denote by $\chi_{X}^{\alpha\alpha^{\prime}}(\bm{q},i\omega_{n})$ the susceptibility obtained from Eq.~\eqref{eq:matsubara-spin-susceptibility} with $\chi_{ij}^{\alpha\alpha^{\prime}}(\tau)$ replaced by $\chi_{ij,X}^{\alpha\alpha^{\prime}}(\tau)$, where $X=\mathrm{I},\mathrm{II}$.
The component-resolved dynamical spin structure factor for each decoupling scheme is then obtained as
\begin{align}
    \label{eq:dynamical-ssf-from-matsubara}
    S_{X}^{\alpha\alpha^{\prime}}(\bm{q},\omega)
    =
    \frac{2}{1-e^{-\beta\omega}}
    \mathrm{Im}\,
    \chi_{X}^{\alpha\alpha^{\prime}}(\bm{q},\omega+i\delta),
    \qquad
    X=\mathrm{I},\mathrm{II}.
\end{align}
The equal-time structure factor is obtained from Eq.~\eqref{eq:static-spin-structure-factor-matsubara}.
The distinction between decoupling I and decoupling II therefore enters before analytic continuation, at the level of the Matsubara susceptibility itself.
Decoupling I evaluates the spin-spin correlator as a spinon bubble, whereas decoupling II evaluates it as a bond-channel bubble consistent with the SBMFT saddle point.
Because decoupling I and decoupling II retain different sets of two-point contractions, they need not give equivalent spin correlations for bond-dependent interactions.
Below, we examine this difference explicitly in the Kitaev model and show how the bond-operator decoupling reflects the bond-dependent anisotropy of the interaction in the spin structure factors.

\section{Model}
\label{sec:Model}

\begin{figure}[t]
  \centering
      \includegraphics[width=\columnwidth,clip]{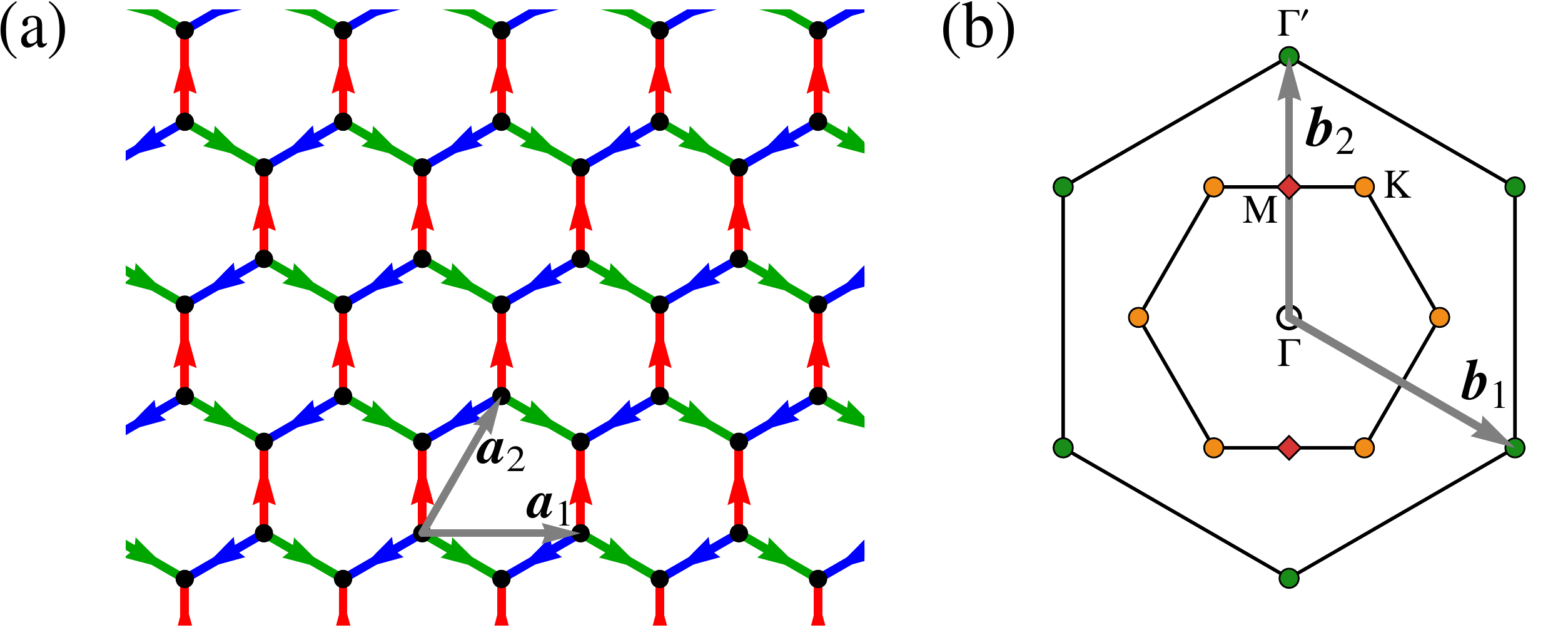}
      \caption{
        (a)
        Schematic picture of the honeycomb lattice on which the $S=1$ Kitaev model is defined.
        The blue, green, and red lines denote the $x$, $y$, and $z$ bonds, respectively.
        The gray arrows indicate the primitive translation vectors $\bm{a}_{1}$ and $\bm{a}_{2}$ of the two-sublattice honeycomb unit cell.
        For a directed $\gamma$ bond $(i,j)$ with $\gamma=x,y,z$, we introduce the mean-field amplitudes
        $\left(
        \braket{\mathcal{A}_{ij}},
        \braket{\mathcal{B}_{ij}},
        \braket{\mathcal{C}^{\gamma}_{ij}},
        \braket{\mathcal{D}^{\gamma}_{ij}}
        \right)$
        from site $i$ to site $j$.
        The arrows on the bonds specify this direction convention.
        The mean-field amplitudes are allowed to depend on the bond type $\gamma$, namely on whether the bond is an $x$, $y$, or $z$ bond, while they are taken to be translationally invariant with respect to the primitive two-sublattice unit cell.
        (b)
        First and extended Brillouin zones of the original honeycomb lattice.
        Filled symbols denote the high-symmetry points.
        The gray arrows indicate the primitive reciprocal lattice vectors $\bm{b}_{1}$ and $\bm{b}_{2}$ corresponding to $\bm{a}_{1}$ and $\bm{a}_{2}$ in real space.
      }
      \label{fig:mean_field_ansatz}
\end{figure}

We apply the SBMFT, as explained in the previous section, to the $S=1$ Kitaev model on a honeycomb lattice, whose elementary excitations were predicted to be bosonic in a previous study~\cite{Ma-2023}.
The Hamiltonian of this model is given by 
\begin{align}
    \label{eq:Kitaev-Hamiltonian}
    \mathcal{H}=K \sum_{\gamma=x,y,z} \sum_{\langle i,j\rangle_{\gamma}}S_{i}^{\gamma}S_{j}^{\gamma},
\end{align}
where $S_{i}^{\gamma} \left(\gamma=x,y,z\right)$ denotes the $S=1$ spin operator at site $i$, and $K$ is the exchange constant of the Kitaev interaction between spins on nearest-neighbor sites.
The Kitaev interaction is bond-dependent, and $\langle i,j\rangle_{\gamma}$ denotes the nearest-neighbor $\gamma$ bond on the honeycomb lattice [see Fig.~\ref{fig:mean_field_ansatz}(a)].
The sign of $K$ determines the magnetic nature of the model: $K>0$ corresponds to the antiferromagnetic (AFM) Kitaev model, whereas $K<0$ corresponds to the ferromagnetic (FM) Kitaev model.
Throughout this paper, we measure energies in units of $\left|K\right|$ and set the length of the primitive translation vectors of the honeycomb lattice to unity.

Figure~\ref{fig:mean_field_ansatz}(b) shows the first and extended Brillouin zones of the original honeycomb lattice together with the conventional high-symmetry points.
The main-text calculations for the AFM Kitaev model are performed using the primitive two-sublattice unit cell shown in Fig.~\ref{fig:mean_field_ansatz}(a).
The zero-flux spin-triplet ansatz introduced in Sec.~\ref{sec:Mean-field ansatz} is translationally invariant with respect to this primitive unit cell; the three colors in Fig.~\ref{fig:mean_field_ansatz}(a) label only the $x$, $y$, and $z$ Kitaev bond types.
The FM and AFM Kitaev models are related by an exact sublattice-dependent $\pi$ rotation of the spin quantization axes.
The details are given in Appendix~\ref{app:four-sublattice transformation of the Kitaev model}.
\section{Results}
\label{sec:results}

In this section, we present our numerical results for the $S=1$ Kitaev model.  
First, we show the zero-temperature results for the dynamical spin structure factor $S(\bm{q},\omega)$ given in Eq.~\eqref{eq:dynamical spin structure factor}, and the equal-time spin structure factor $S(\bm{q})$ in Eq.~\eqref{eq:static spin structure factor} in Sec.~\ref{sec:Zero-temperature}.  
Next, in Sec.~\ref{sec:Finite-temperature}, we extend the study to finite temperatures, primarily focusing on the dynamical spin structure factor $S(\bm{q},\omega)$.  
In all cases, the spin structure factors are presented as the trace over the three spin components,  
\begin{align}
    \label{eq:trace spin structure factor}
    S(\bm{q},\omega)
    &=\frac{1}{3}\sum_{\alpha=x,y,z}S^{\alpha\alpha}(\bm{q},\omega),\\
    S(\bm{q})
    &=\frac{1}{3}\sum_{\alpha=x,y,z}S^{\alpha\alpha}(\bm{q}).
\end{align}
In the numerical evaluation of the dynamical spin structure factor, we use the broadening parameter $\delta/|K|=0.1$.

\subsection{Mean-field ansatz}
\label{sec:Mean-field ansatz}

\begin{figure}[t]
  \centering
      \includegraphics[width=\columnwidth,clip]{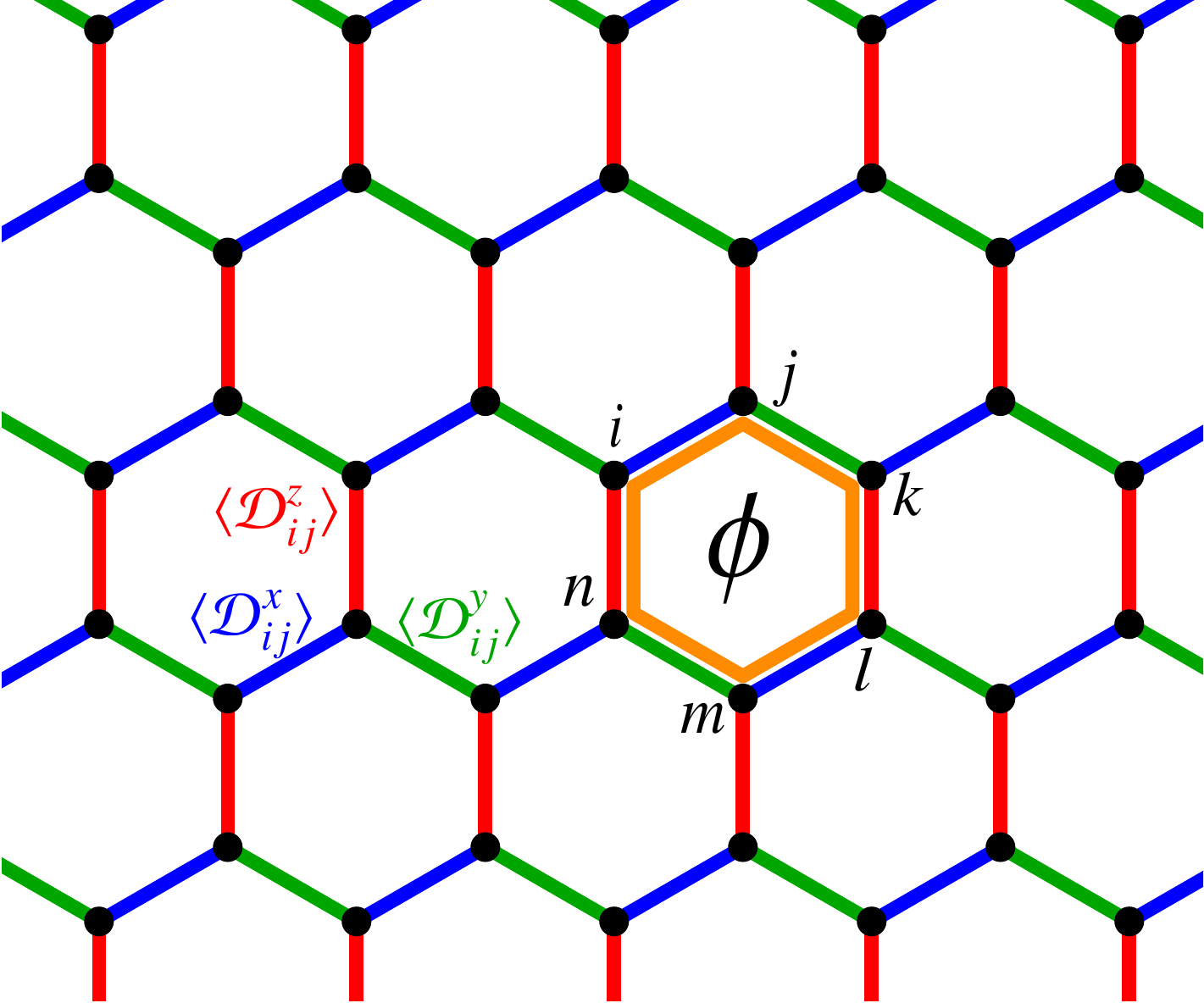}
      \caption{
      Mean-field pattern of the zero-flux spin-triplet ansatz $\phi_{t}=0$ adopted in the present SBMFT for the $S=1$ AFM Kitaev model.
      On each $\gamma$ bond, the only nonzero intersite mean-field amplitude is $\braket{\mathcal{D}_{ij}^{\gamma}}=D$, with a common real value for $\gamma=x,y,z$.
      Since $\mathcal{D}_{ij}^{\gamma}=\mathcal{D}_{ji}^{\gamma}$, the mean fields are independent of bond orientation; accordingly, the arrows used in Fig.~\ref{fig:mean_field_ansatz}(a) are omitted here.
      }
      \label{fig:mean_field_result}
\end{figure}

Before presenting the spin structure factors, we briefly summarize the mean-field ansatz employed in this study.  
By performing the mean-field approximation given in Eq.~\eqref{eq:local Hamiltonian} to the Hamiltonian of the $S=1$ Kitaev model, the Kitaev interaction on the $\gamma$ bond $\langle i,j\rangle_\gamma$ is expressed in terms of four mean-field parameters: $\braket{\mathcal{A}_{ij}}$, $\braket{\mathcal{B}_{ij}}$, $\braket{\mathcal{C}_{ij}^\gamma}$, and $\braket{\mathcal{D}_{ij}^\gamma}$, where the direction from $i$ to $j$ is depicted in Fig.~\ref{fig:mean_field_ansatz}(a).
The mean-field pattern used in the main-text calculation is shown in Fig.~\ref{fig:mean_field_result}.
We label its flux sector by the Wilson-loop phase associated with the anomalous-pairing amplitudes around an elementary hexagon.
For the pattern in Fig.~\ref{fig:mean_field_result}, this phase is zero, and we therefore denote the ansatz by $\phi_t=0$.
The explicit convention for the spin-singlet and spin-triplet Wilson-loop phases, $\phi_s$ and $\phi_t$, is given in Appendix~\ref{app:spin-structure-factors-in-other-flux-sectors}.
In the main-text calculations, we adopt the zero-flux spin-triplet ansatz of Ref.~\cite{Ralko-Merino-2024}.
In this ansatz, the nonzero intersite mean-field amplitude is
\begin{align}
    \label{eq:zero-flux-triplet-d-ansatz}
    \braket{\mathcal{D}_{ij}^{\gamma}} = D
    \qquad
    \text{on a $\gamma$ bond},
\end{align}
with a common real amplitude $D$ for $\gamma=x,y,z$, while
\begin{align}
    \label{eq:zero-flux-triplet-vanishing-intersite-fields}
    \braket{\mathcal{A}_{ij}}
    =
    \braket{\mathcal{B}_{ij}}
    =
    \braket{\mathcal{C}_{ij}^{\gamma}}
    =
    0
    \qquad
    \text{for intersite bonds}.
\end{align}
The amplitude $D$ and the Lagrange multiplier are determined self-consistently within this ansatz.
With the Wilson-loop convention defined in Appendix~\ref{app:spin-structure-factors-in-other-flux-sectors}, this pattern gives $e^{i\phi_t}=1$, and hence belongs to the zero-flux triplet sector $\phi_t=0$ in the notation of Ref.~\cite{Ralko-Merino-2024}.

We now clarify why we focus on the zero-flux spin-triplet ansatz, $\phi_t=0$, in the main text. We first emphasize that this choice is not intended to supersede the energetic and
ED-based selection discussed in Ref.~\cite{Ralko-Merino-2024}.
In Ref.~\cite{Ralko-Merino-2024}, several competing SBMFT saddle points were compared in terms of their mean-field energies and their spin excitation spectra, and a chiral spin-liquid state was proposed as a promising candidate for the integer-spin AFM Kitaev model.
In the present work, our aim is complementary: we investigate the spin dynamics of the $S=1$ Kitaev model within a time-reversal-symmetric mean-field description that explicitly retains the bond-dependent spin anisotropy of the Kitaev interaction.
The reason for considering a spin-triplet ansatz is that the pure Kitaev interaction couples the spin component to the bond direction as $S_{i}^{\gamma}S_{j}^{\gamma}$.
The spin-singlet flux sectors discussed in Ref.~\cite{Ralko-Merino-2024}, $\phi_s=0,\pi,\pi/2$, are legitimate SBMFT saddle points in their corresponding stability regimes.
However, in these states the finite intersite pairing amplitude belongs to the SU($2$)-invariant singlet channel, while the SU(2)-breaking triplet amplitudes $\langle \mathcal{D}_{ij}^{x,y,z}\rangle$ vanish.
Consequently, the mean-field Hamiltonian treats the three spin components on the same footing at the level of the nonzero pairing channels.
This feature is directly reflected in the component-resolved equal-time structure factors shown in Fig.~\ref{fig:other_singlet_ansatz_SSF}: for the spin-singlet ansatze, $S^{xx}(\bm{q})$, $S^{yy}(\bm{q})$, and $S^{zz}(\bm{q})$ show identical momentum dependence.
This behavior is not a consequence of the particular decoupling scheme used to evaluate spin correlations, but follows from the SU($2$)-invariant structure of the singlet mean-field ansatz itself.

By contrast, a spin-triplet ansatz with finite $\langle\mathcal{D}_{ij}^{\gamma}\rangle$ on a $\gamma$ bond encodes the bond-spin locking of the Kitaev interaction already at the mean-field level. 
Among the triplet flux sectors discussed in Ref.~\cite{Ralko-Merino-2024}, the $\phi_t=\pi$ state loses its spin-liquid stability before reaching the $S=1$ regime, whereas the $\phi_t=\pi/2$ state is a chiral state that breaks time-reversal symmetry.
Since the higher-spin Kitaev picture based on composite bosonic excitations coupled to a bond gauge field does not require time-reversal-symmetry breaking, we focus in the main text on the remaining time-reversal-symmetric triplet sector, namely the zero-flux state $\phi_t=0$.
We therefore use this ansatz as a representative time-reversal-symmetric spin-triplet Kitaev spin-liquid state and analyze its zero- and finite-temperature spin dynamics.
We do not claim that this ansatz is the lowest-energy SBMFT saddle point of the AFM $S=1$ Kitaev model.
Thus, our use of this ansatz should be understood as a comparison of spin-correlation evaluation schemes within a fixed mean-field background, rather than as an energetic determination of the ground-state ansatz.

For completeness, and to facilitate comparison with the competing flux sectors discussed in Ref.~\cite{Ralko-Merino-2024}, we also present the zero-temperature dynamical spin structure factors and component-resolved equal-time structure factors for the other flux sectors that remain spin-liquid solutions at $S=1$ within the present SBMFT calculation, namely $\phi_t=\pi/2$, $\phi_s=\pi$, and $\phi_s=\pi/2$, in Appendix~\ref{app:spin-structure-factors-in-other-flux-sectors}.
The $\phi_t=\pi$ and $\phi_s=0$ sectors are not included in this comparison because they do not remain spin-liquid solutions up to $S=1$.
For the FM Kitaev model, the mean-field ansatz is obtained by applying the sublattice-rotation mapping introduced in Sec.~\ref{sec:Model} and Appendix~\ref{app:four-sublattice transformation of the Kitaev model} to the AFM Kitaev model.
As a numerical check, we also performed the self-consistent SBMFT calculation for the FM Hamiltonian itself, using the transformed AFM ansatz as the initial mean-field configuration, and confirmed that the mean-field parameters converge to the corresponding transformed solution.

Before presenting the spin structure factors, we comment on the
on-site contribution to the equal-time spin correlations. In an
exact treatment, the equal-time structure factor satisfies the sum
rule
\begin{align}
    \label{eq:sum rule}
    \frac{M}{N}\sum_{\bm q}^{\rm B.Z.} S(\bm q)
    =
    \frac{1}{3}\sum_{\gamma}
    \frac{1}{N}\sum_i
    \braket{S_i^\gamma S_i^\gamma}
    =
    \frac{S(S+1)}{3},
\end{align}
which follows from the fixed spin length on each site. For $S=1$,
the right-hand side is $2/3$. In SBMFT, however, the local
constraint is imposed only on average, and therefore the exact
on-site identity is not generally preserved. The resulting violation
of the spin sum rule in the conventional SBMFT
calculation has been discussed previously, for example in
Ref.~\cite{Mezio-2011}.
We now evaluate the corresponding on-site contribution within
decoupling II for the Ising-interaction representation used in the
present calculation. For this purpose, we evaluated the four on-site
bond-operator channels appearing in the Ising representation.
We found that only the $\mathcal{B}_{ii}$ channel contributes, while the
$\mathcal{A}_{ii}$, $\mathcal{C}_{ii}^{\gamma}$, and
$\mathcal{D}_{ii}^{\gamma}$ channels vanish within numerical accuracy.
Since $\mathcal{B}_{ii}=n_i/2$, the on-site part of the momentum sum in
decoupling II is determined by the local boson-number contribution.
For the Ising bond-operator representation used in the present
calculation, this gives
\begin{align}
    \label{eq:momentum-sum-rule-decoupling-II}
    \frac{M}{N}
    \sum_{\bm{q}}^{\mathrm{B.Z.}}
    S_{\mathrm{II}}(\bm{q})
    =
    \frac{S(S+1)}{2}.
\end{align}
For $S=1$, Eq.~\eqref{eq:momentum-sum-rule-decoupling-II} gives $1$,
whereas the exact spin sum rule in Eq.~\eqref{eq:sum rule} gives $2/3$.
Thus, decoupling II still violates the exact spin sum rule and gives a
value larger by a factor of $3/2$ for $S=1$, in the same sense as the sum-rule violation of decoupling I discussed in Ref.~\cite{Mezio-2011}.
We have numerically verified that the momentum sum of the calculated
equal-time structure factor satisfies Eq.~\eqref{eq:momentum-sum-rule-decoupling-II}
within numerical accuracy.

\subsection{Zero-temperature spin structure factor}
\label{sec:Zero-temperature}

\begin{figure*}[t]
  \centering
      \includegraphics[width=2\columnwidth,clip]{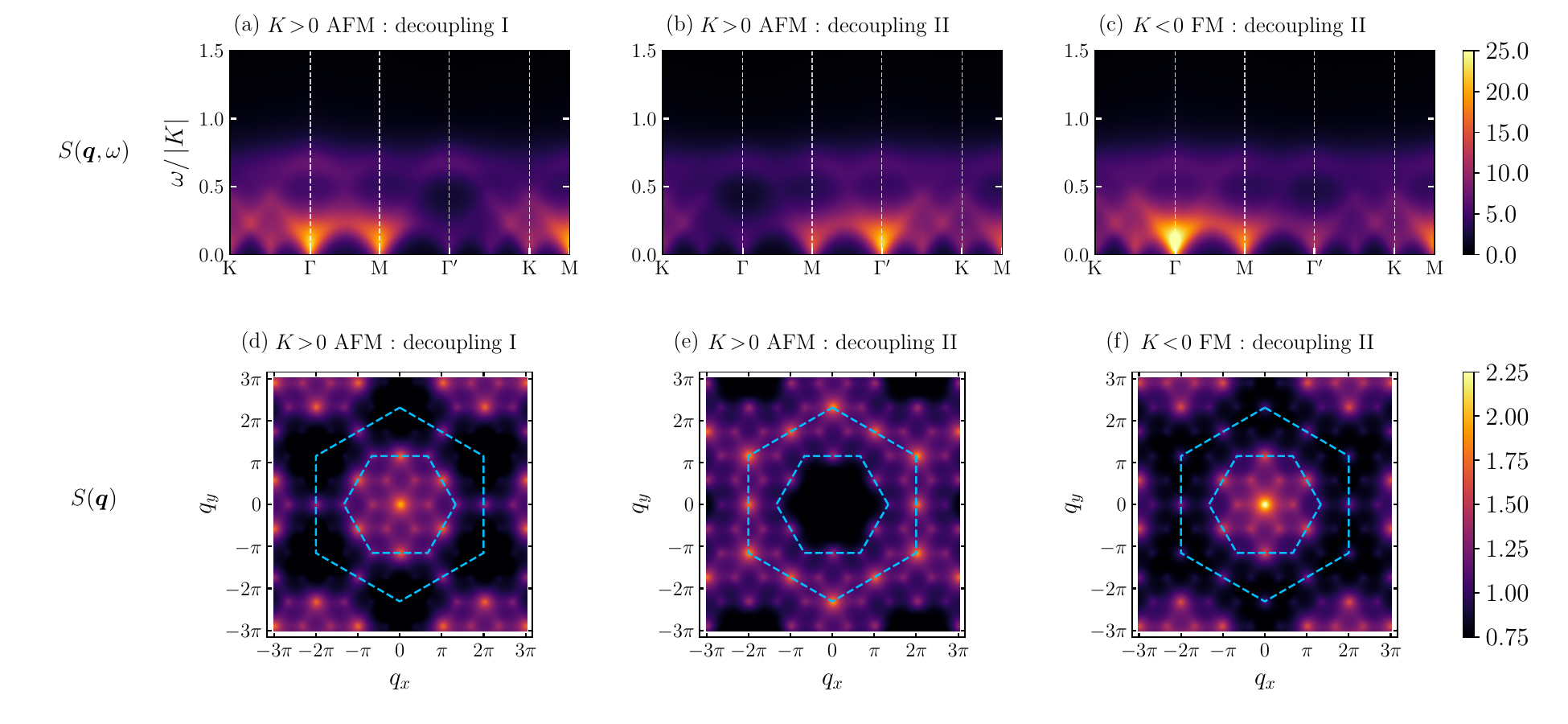}
      \caption{
        Dynamical (upper row) and equal-time (lower row) spin structure factors in the ground state of the $S=1$ Kitaev model.  
        In (a)--(c), the spectra are shown along the path connecting the high-symmetry points indicated in Fig.~\ref{fig:mean_field_ansatz}.  
        In (d)--(f), the inner and outer hexagons represent the first and extended Brillouin zones, respectively.  
        Panels (a) and (d) correspond to results obtained using decoupling I in the AFM model ($K>0$), panels (b) and (e) to results obtained using decoupling II in the AFM model, and panels (c) and (f) to results obtained using decoupling II in the FM model ($K<0$).  
      }
      \label{fig:ground_state}
\end{figure*}

Here, we present zero-temperature results for the dynamical and equal-time spin structure factors in the $S=1$ Kitaev model.
In Sec.~\ref{sec:Calculation of the spin structure factor}, we introduced two inequivalent decoupling schemes for the spin-spin correlator, referred to as decoupling I and decoupling II.
Note that decoupling I was applied to the AFM Kitaev model in previous work~\cite{Ralko-Merino-2024}, whereas decoupling II is a new approach introduced in this paper.

Figures~\ref{fig:ground_state}(a) and \ref{fig:ground_state}(d) present the dynamical and equal-time spin structure factors obtained using decoupling I, respectively.  
We observe that the dynamical spin structure factor $S(\bm{q},\omega)$ exhibits pronounced low-energy weight around the $\Gamma$ and $\mathrm{M}$ points, while the spectral weight at the $\Gamma^{\prime}$ point is suppressed.  
Such features are also observed in the equal-time spin structure factor $S(\bm{q})$, which shows a strong peak at the $\Gamma$ point, with the signal at the $\Gamma^{\prime}$ point being essentially absent.  
The dynamical and equal-time spin structure factors obtained here using decoupling I are consistent with the results reported in Ref.~\cite{Ralko-Merino-2024}.  
However, for the present nearest-neighbor AFM Kitaev ansatz, this momentum dependence of the equal-time spin structure factor is not consistent with the sign structure suggested by the dominant bond correlations: the frequency-integrated weight is expected to be enhanced around the $\Gamma^{\prime}$ points rather than dominated by the $\Gamma$ point.

To resolve this issue, we now turn to the results obtained using decoupling II, as introduced in Sec.~\ref{sec:Calculation of the spin structure factor}.
Figures~\ref{fig:ground_state}(b) and \ref{fig:ground_state}(e) show the results obtained with decoupling II for the dynamical and equal-time spin structure factors, respectively.
We find contrasting behavior when applying decoupling II: the intensity at the $\Gamma$ point is strongly suppressed, while a large spectral weight is observed at the $\Gamma^{\prime}$ point, consistent with N\'eel-type correlations.
The $S=1/2$ AFM Kitaev model exhibits similar features in the dynamical spin structure factor~\cite{
Knolle-2014,
Knolle-2015,
Yoshitake-Nasu-Motome-2016,
Yoshitake-Nasu-2017,
Laurell-2020,
Nasu-Motome-2021,
Takegami-2025}.
We note that a finite-frequency response at the $\Gamma$ point can indeed be present in the AFM Kitaev model, as also pointed out in previous studies.
Nevertheless, when the dynamical response is viewed over the full energy range, the frequency-integrated spectral weight reflects the AFM sign structure of the Kitaev interaction, exhibiting a stronger weight around the $\Gamma^{\prime}$ points than around the $\Gamma$ point.
From this perspective, decoupling II, which preserves the bond-dependent structure of the Kitaev interaction in the evaluation of spin correlations, provides a more suitable scheme for computing the dynamical spin structure factor than decoupling I in the present ansatz.

Our findings clearly demonstrate that the choice of decoupling scheme significantly affects the resulting spin structure factor, with decoupling II yielding 
results consistent with the AFM correlations in the present ansatz of the AFM Kitaev model.
We also comment on why such a strong decoupling-scheme dependence does not arise for the spin-singlet ansatze.
As discussed in Appendix~\ref{app:spin-structure-factors-in-other-flux-sectors}, these ansatze contain only the SU($2$)-invariant singlet-pairing channel $\mathcal{A}_{ij}$ as a finite intersite mean field.
Consequently, the component-resolved spin correlations satisfy $S^{xx}(\bm{q})=S^{yy}(\bm{q})=S^{zz}(\bm{q})$, and the results obtained using decoupling I and decoupling II coincide.
This explains why previous SBMFT studies employing the conventional scheme, i.e., decoupling I~\cite{
Lefmann-Hedegard-1994,
Mezio-2011,
Merino-Holt-Powell-2014,
Halimeh-Punk-2016,
Bauer-2017,Mondal-2017,
Halimeh-Singh-2019,
Mondal-2021,
Rossi-Motruk-2023}, are appropriate for discussing spin correlations in Heisenberg models, whereas the choice of decoupling scheme becomes crucial for bond-dependent anisotropic interactions such as the Kitaev interaction.

We also present the results for the FM Kitaev model ($K<0$) in Figs.~\ref{fig:ground_state}(c) and \ref{fig:ground_state}(f), which are obtained by applying the sublattice-rotation mapping introduced in Sec.~\ref{sec:Model} to the AFM mean-field data.
We have verified this correspondence by solving the FM model directly within the same self-consistent mean-field scheme and confirming convergence of the transformed FM mean-field solution.
As expected for FM interactions, the dynamical spin structure factor $S(\bm{q},\omega)$ exhibits a pronounced low-energy peak at the $\Gamma$ point, whereas the spectral weight at the $\Gamma^{\prime}$ point is significantly suppressed.  
These characteristics are also evident in the equal-time spin structure factor $S(\bm{q})$, where a strong peak appears at the $\Gamma$ point and the signal at the $\Gamma^{\prime}$ point is notably weak.  
These findings further support the usefulness
of decoupling II for analyzing the present Kitaev mean-field ansatz.
A detailed discussion regarding the choice of the decoupling scheme used to compute the spin structure factor is provided in Sec.~\ref{sec:Discussion}.

\subsection{Finite-temperature dynamical spin structure factor}
\label{sec:Finite-temperature}

\begin{figure*}[t]
  \centering
      \includegraphics[width=2\columnwidth,clip]{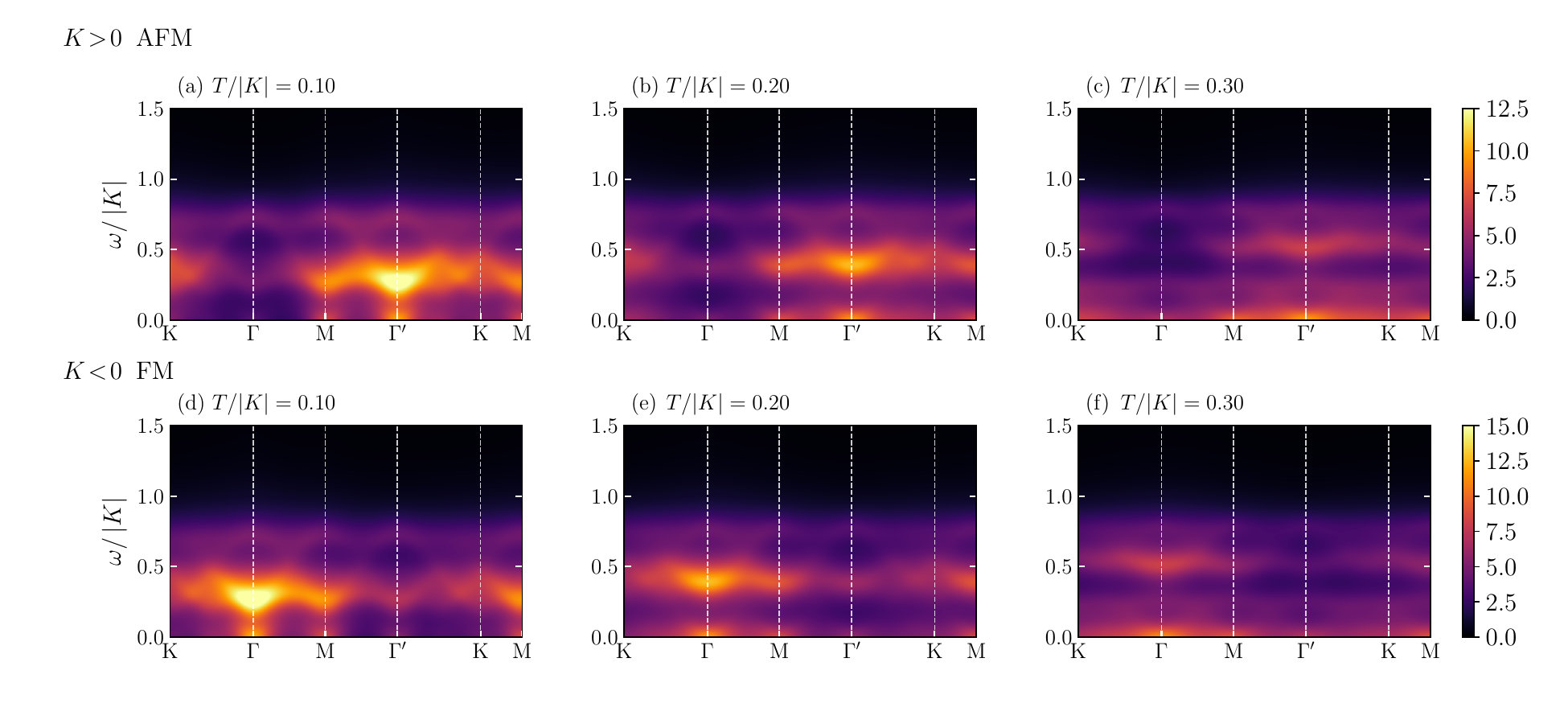}
      \caption{Finite-temperature dynamical spin structure factor $S(\bm{q},\omega)$ in the AFM and FM Kitaev models at (a),(d) $T=0.1$, (b),(e) $T=0.2$, and (c),(f) $T=0.3$.
       Panels (a)--(c) are the results in the AFM model ($K>0$), whereas panels (d)--(f) are for the FM model ($K<0$).}
      \label{fig:finite_temperature}
\end{figure*}

\begin{figure}[t]
  \centering
      \includegraphics[width=\columnwidth,clip]{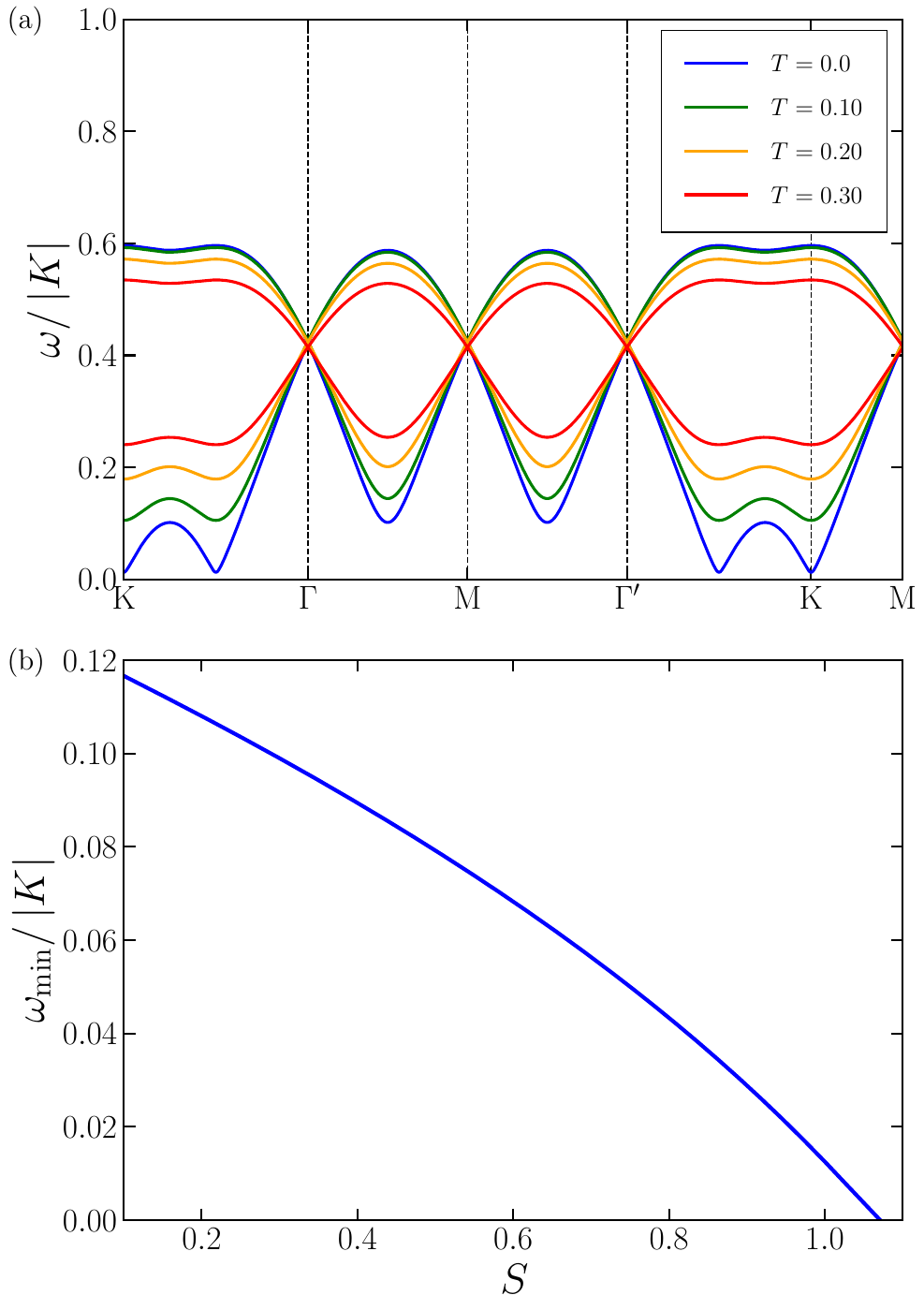}
      \caption{
        (a) Spinon dispersions in the $S=1$ AFM Kitaev model along the path connecting the high-symmetry points indicated in Fig.~\ref{fig:mean_field_ansatz}.
        At $T=0$, the spectrum is nearly gapless, but a small gap remains.
        (b) Zero-temperature excitation gap $\omega_{\text{min}}/|K|$ of spinons as a function of the spin quantum number $S$.}
      \label{fig:dispersion}
\end{figure}

\begin{figure}[t]
  \centering
      \includegraphics[width=\columnwidth,clip]{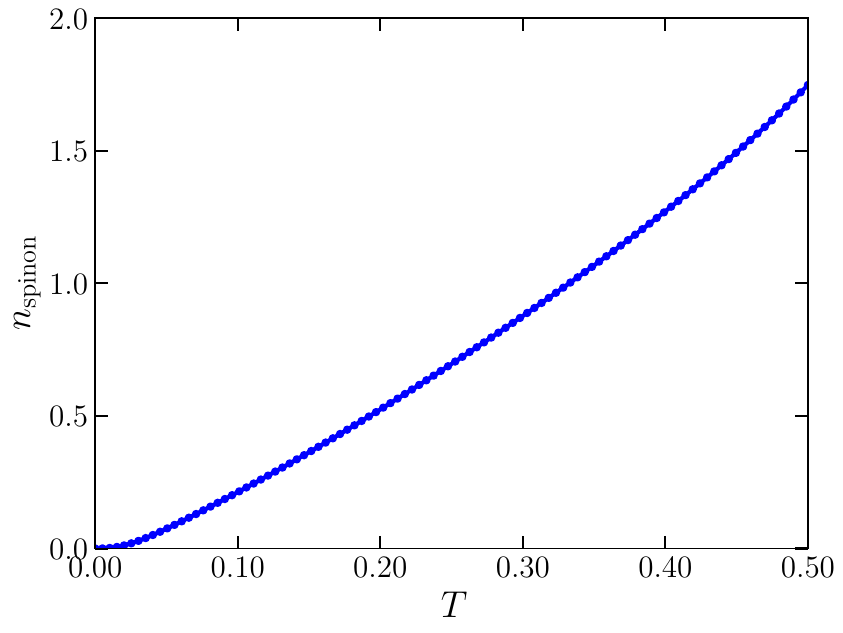}
      \caption{Spinon number per site, $n_{\rm spinon}$, in the $S=1$ AFM Kitaev model as a function of temperature calculated by SBMFT.
      }
      \label{fig:spinon_density}
\end{figure}

We now turn to the temperature evolution of the dynamical spin structure factor $S(\bm{q},\omega)$ for the $S=1$ Kitaev model using decoupling II.
It is well known that evaluating finite-temperature dynamics in QSLs is challenging due to the sign problem in frustrated magnets, which limits the applicability of quantum Monte Carlo simulations.
While the $S=1/2$ Kitaev model has been studied at finite temperatures by exploiting its exact solvability, the $S=1$ case remains largely unexplored as it is not exactly solvable.
Here, we adopt the SBMFT, which enables us to compute the finite-temperature dynamical spin structure factor without encountering the sign problem.

We present the temperature evolution of the dynamical spin structure factor $S(\bm{q},\omega)$ for the $S=1$ Kitaev models in Fig.~\ref{fig:finite_temperature}.
Figures~\ref{fig:finite_temperature}(a)--\ref{fig:finite_temperature}(c) show $S(\bm{q},\omega)$ at $T=0.10$, $0.20$, and $0.30$ for the AFM Kitaev model ($K>0$), while Figs.~\ref{fig:finite_temperature}(d)--\ref{fig:finite_temperature}(f) display the corresponding results for the FM Kitaev model ($K<0$).
First, we focus on the AFM case.
At zero temperature, $S(\bm{q},\omega)$ exhibits a continuum with significant spectral weight around the $\Gamma^{\prime}$ point, appearing as a broad peak from $\omega=0$ to $\omega/|K|\approx 0.3$, as shown in Fig.~\ref{fig:ground_state}(b).
As the temperature increases, the spectral weight at the $\Gamma^{\prime}$ point gradually shifts to higher energies and appears to split into two peaks around $\omega/|K|=0.05$ and $\omega/|K|=0.3$, as shown in Figs.~\ref{fig:finite_temperature}(a)--\ref{fig:finite_temperature}(c).
We find that the high-energy weight around $\omega/|K|\approx 0.3$ 
decreases as the temperature increases, accompanied by a shift to higher energies.
On the other hand, the low-energy weight around $\omega/|K|\approx 0.05$ is not significantly suppressed, and its peak position remains nearly unchanged with increasing temperature, in contrast to the high-energy weight.
Moreover, the low-energy peak around the $\Gamma^{\prime}$ point broadens in momentum $\bm{q}$, indicating a loss of coherence in the spinon excitations associated with the N\'eel order as the temperature rises.

Figures~\ref{fig:finite_temperature}(d)--\ref{fig:finite_temperature}(f) show the temperature evolution for the FM counterpart ($K<0$).  
The zero-temperature spectrum is presented in Fig.~\ref{fig:ground_state}(c).  
We find that the trend in temperature evolution is similar to that of the AFM case, except for the $\bm{q}$ dependence, since it is derived from the AFM data through the sublattice-dependent transformation introduced in Sec.~\ref{sec:Model}: the spectral weight around the $\Gamma$ point is pronounced at zero temperature, and it shifts to higher energies and splits into two distinct structures as the temperature increases.
For both the AFM and FM models, we observe that the spectral weight in $S(\bm{q},\omega)$ gradually diminishes with increasing temperature.

The features in the temperature evolution of $S(\bm{q},\omega)$ presented above reflect the temperature dependence of the underlying spinon in Fig.~\ref{fig:dispersion}(a).
Within this framework, the spinons are free bosons whose dispersion acquires a gap that increases with temperature as shown in Fig.~\ref{fig:dispersion}(a).
Since the instability to a long-range magnetic order is associated with the vanishing of this gap, raising temperature suppresses such ordering tendencies and correspondingly it shifts spectral weight to higher energies.
As the gap increases, the spinon bandwidth narrows; this real-space localization manifests in momentum space as dispersionless structure in the dynamical structure factor, as described above.
Simultaneously, the decrease in intensity arises from enhanced thermal fluctuations, which appear in the dynamical structure factor through the Bose-Einstein distribution.
The similarity in the temperature dependences of the FM and AFM Kitaev models reflects their relation via a unitary transformation.

Finally, we assess the validity of the noninteracting Schwinger boson description at finite temperatures.
This treatment is justified as long as the density of thermally excited spinons remains low, which is evaluated by the average number of spinons per site, $n_{\text{spinon}}(T)$, defined as
\begin{align}
    \label{eq:spinon number}
    n_{\text{spinon}}(T)&=\frac{1}{N}\sum_{\bm{k}}\sum_{n=1}^{2M}\frac{1}{e^{\varepsilon_{\bm{k},n}/T}-1},
\end{align}
where the summation over $n$ runs through the $2M$ spinon branches with positive energy $\varepsilon_{\bm{k},n}$.
For the $S=1$ Kitaev model, the local constraint in Eq.~\eqref{eq:local constraint} imposes an upper limit on the number of spinons per site, $n_{\text{max}}=2S=2$.
Figure~\ref{fig:spinon_density} displays $n_{\text{spinon}}(T)$ as a function of temperature.
Even at the highest temperature ($T=0.30$) examined in the dynamical spin structure factor in Fig.~\ref{fig:finite_temperature}, the average number of spinons per site remains sufficiently small compared to the upper bound $n_{\text{max}}=2$.
These results indicate that the noninteracting Schwinger boson description remains valid at least up to $T=0.30$, and the interaction effects between spinons are expected to play only a minor role.
Therefore, the present finite-temperature analysis is reliable for the dynamical spin structure factor $S(\bm{q},\omega)$ presented in Fig.~\ref{fig:finite_temperature}.

\section{Discussion}
\label{sec:Discussion}
\begin{figure}[t]
  \centering
      \includegraphics[width=\columnwidth,clip]{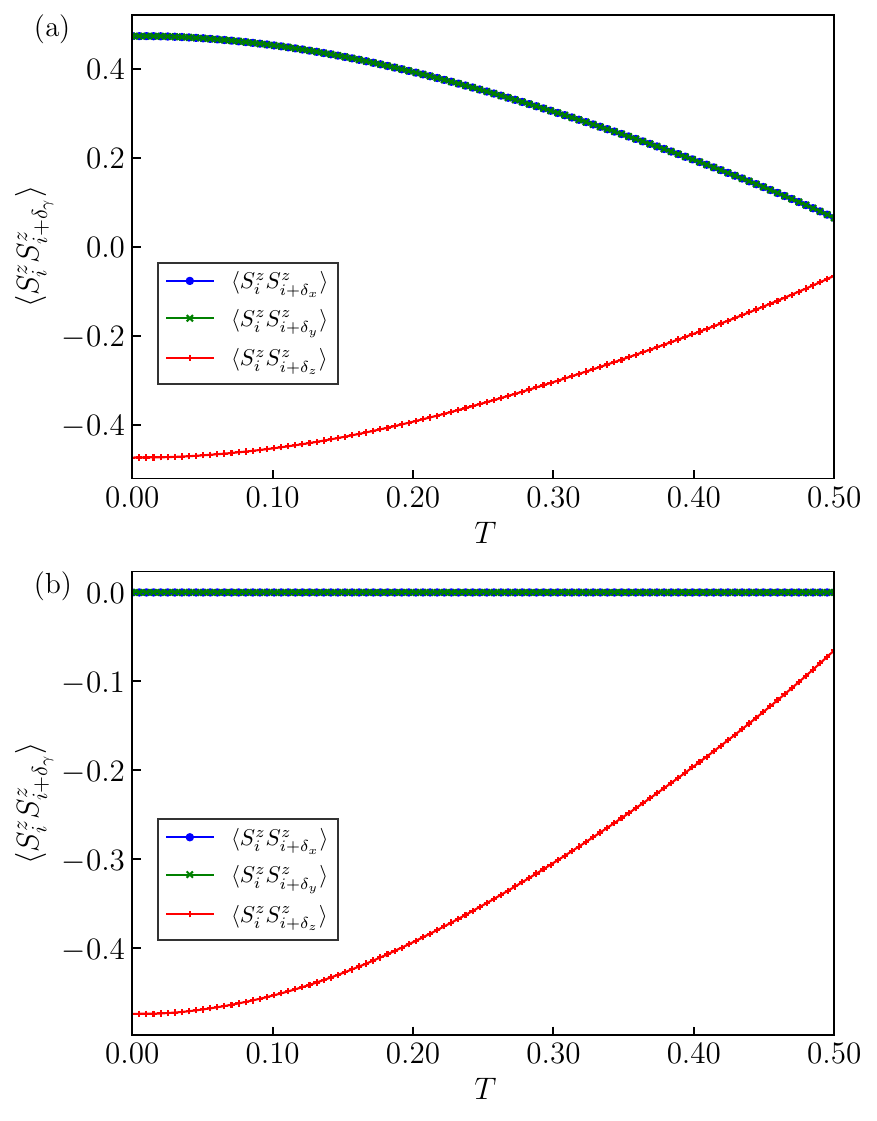}
      \caption{
        Temperature dependence of the spin correlations $\braket{S_{i}^{z}S_{i+\delta_{\gamma}}^{z}}$ between sites $i$ and $i+\delta_\gamma$, where $\delta_\gamma$ denotes the vector connecting the two sites on the $\gamma$ bond, in the $S=1$ AFM Kitaev model, obtained using SBMFT with (a) decoupling I and (b) decoupling II.
        In both figures, the plots of $\braket{S_{i}^{z}S_{i+\delta_{x}}^{z}}$ and $\braket{S_{i}^{z}S_{i+\delta_{y}}^{z}}$ completely overlap.
      }
      \label{fig:spin_correlator}
\end{figure}

In this section, we discuss the inequivalence of the two decoupling schemes introduced in Sec.~\ref{sec:Calculation of the spin structure factor} by focusing on the $S=1$ AFM Kitaev model.
As mentioned in Sec.~\ref{sec:Calculation of the spin structure factor}, the two decoupling schemes yield different results for the spin structure factor.
In decoupling I, the spin structure factor exhibits a peak at the $\Gamma$ point, as shown in Fig.~\ref{fig:ground_state}(d), despite the AFM nature of the Kitaev model, suggesting that this decoupling scheme yields FM-like correlations that are not consistent with the AFM exchange sign structure.
By contrast, decoupling II produces a peak at the $\Gamma^{\prime}$ point, indicating AFM correlations, and no peak at the $\Gamma$ point, as shown in Fig.~\ref{fig:ground_state}(e), results that are consistent with the AFM interaction.

To clarify the origin of the difference between the two decoupling schemes, we analyze spin correlations on nearest-neighbor bonds in the $S=1$ AFM Kitaev model.  
Figures~\ref{fig:spin_correlator}(a) and \ref{fig:spin_correlator}(b) show the temperature dependence of $\braket{S_{i}^{z}S_{j}^{z}}$ on the nearest-neighbor $\gamma$ bond ($\gamma=x,y,z$) using the SBMFT with decoupling I and decoupling II, respectively.  
In decoupling I, the spin correlation $\braket{S_{i}^{z}S_{j}^{z}}$ on the $z$ bond exhibits a negative value at low temperatures, indicating the expected AFM correlation.  
However, the spin correlations on the $x$ and $y$ bonds are nonzero and positive, suggesting an FM-like contribution that is not expected from the nearest-neighbor AFM Kitaev exchange. 
More crucially, the spin correlations on the $x$ and $y$ bonds are zero in the exact diagonalization, which is guaranteed by the presence of a local conserved quantity on each hexagonal plaquette~\cite{Baskaran-Sen-Shankar-2008}.  
This feature is not captured by the decoupling I scheme leading to nonzero spin correlations on the $x$ and $y$ bonds.
Note that the absolute values of the spin correlations on the three nearest-neighbor bonds appear to be the same as each other.  
Since two of them are FM, the net nearest-neighbor spin correlation $\braket{\bm{S}_{i}\cdot\bm{S}_{j}}$ becomes positive, leading to a $\Gamma$-point peak whose sign structure is not consistent with the nearest-neighbor AFM Kitaev exchange in the static structure factor $S(\bm{q})$.

Figure~\ref{fig:spin_correlator}(b) shows the results obtained using decoupling II.
In this case, the spin correlation $\braket{S_{i}^{z}S_{j}^{z}}$ on the $z$ bond remains negative, and its absolute value in the zero-temperature limit is consistent with the exact-diagonalization result.
Furthermore, the spin correlations on the $x$ and $y$ bonds are essentially zero, indicating that the decoupling II scheme captures the characteristic behavior originating from the presence of the local conserved quantity.
Since the nearest-neighbor spin correlations are negative or zero, the static structure factor $S(\bm{q})$ exhibits a peak at the $\Gamma^{\prime}$ point.
Here, the ground-state energy is evaluated from the expectation value of the original spin Hamiltonian using the nearest-neighbor spin correlations obtained with decoupling II, rather than from the saddle-point energy of the quadratic mean-field Hamiltonian including the Lagrange-multiplier term.
The resulting value, $E_{0}/N\approx -0.71$, is close to the value obtained by exact diagonalization, $E_{0}/N\approx -0.65$, for the $S=1$ Kitaev model~\cite{Koga-2018}.
These observations further support the usefulness of the decoupling II scheme for analyzing the present $S=1$ AFM Kitaev mean-field ansatz.

Here, we clarify the origin of the FM-like correlations that arise in the decoupling I scheme, which lead to a $\Gamma$-point peak that is not consistent with the nearest-neighbor AFM Kitaev exchange in the static spin structure factor.
We focus on the $x$ bond, $\braket{ij}_x$, where the Kitaev-type interaction is represented as $K S_{i}^{x}S_{j}^{x}$.
By applying SBMFT [Eq.~\eqref{eq:local Hamiltonian}] to the bond-operator representation given in Eq.~\eqref{eq:Ising interaction Schwinger boson representation}, the Kitaev-type interaction on the $x$ bond is decoupled as
\begin{align}\label{eq:sxsx_interaction}
    S_{i}^{x}S_{j}^{x}\approx\frac{1}{2}\Bigl(
        &\braket{\mathcal{B}_{ij}^{\dagger}}\mathcal{B}_{ij}
        +\mathcal{B}_{ij}^{\dagger}\braket{\mathcal{B}_{ij}}
        -\braket{\mathcal{B}_{ij}^{\dagger}}\braket{\mathcal{B}_{ij}}\notag\\
        &-\braket{\mathcal{D}_{ij}^{x\dagger}}\mathcal{D}_{ij}^x
        -\mathcal{D}_{ij}^{x\dagger}\braket{\mathcal{D}_{ij}^x}
        +\braket{\mathcal{D}_{ij}^{x\dagger}}\braket{\mathcal{D}_{ij}^x}\notag\\
        &+\braket{\mathcal{C}_{ij}^{x\dagger}}\mathcal{C}_{ij}^x
        +\mathcal{C}_{ij}^{x\dagger}\braket{\mathcal{C}_{ij}^x}
        -\braket{\mathcal{C}_{ij}^{x\dagger}}\braket{\mathcal{C}_{ij}^x}\notag\\
        &-\braket{\mathcal{A}_{ij}^{\dagger}}\mathcal{A}_{ij}
        -\mathcal{A}_{ij}^{\dagger}\braket{\mathcal{A}_{ij}}
        +\braket{\mathcal{A}_{ij}^{\dagger}}\braket{\mathcal{A}_{ij}}
        \Bigr).
\end{align}
The decoupling II scheme follows this procedure, and the expectation value in Eq.~\eqref{eq:sxsx_interaction} can be written as
\begin{align}
    \braket{S_{i}^{x}S_{j}^{x}}\approx\frac{1}{2}\Bigl(
       |\braket{\mathcal{B}_{ij}}|^2
        -|\braket{\mathcal{D}_{ij}^x}|^2
        +|\braket{\mathcal{C}_{ij}^x}|^2
        -|\braket{\mathcal{A}_{ij}}|^2
        \Bigr).
\end{align}
On an $x$ bond, only $\braket{\mathcal{D}_{ij}^{x}}$ is nonzero, and the other expectation values vanish in the mean-field solution, indicating that $\braket{S_{i}^{x}S_{j}^{x}}$ is negative.
In a similar manner, applying the decoupling II scheme to $\braket{S_{i}^{z}S_{j}^{z}}$ yields
\begin{align}
    \braket{S_{i}^{z}S_{j}^{z}}\approx\frac{1}{2}\Bigl(
       |\braket{\mathcal{B}_{ij}}|^2
        -|\braket{\mathcal{D}_{ij}^z}|^2
        +|\braket{\mathcal{C}_{ij}^z}|^2
        -|\braket{\mathcal{A}_{ij}}|^2
        \Bigr),
\end{align}
This is zero on an $x$ bond, since $\braket{\mathcal{D}_{ij}^{x}}$ does not appear in the above representation.

On the other hand, under the decoupling I scheme, the spin correlation $\braket{S_{i}^{z}S_{j}^{z}}$ is decoupled in terms of the original bosonic operators as
\begin{align}
    \label{eq:szsz_decoupling I}
    \braket{S_{i}^{z}S_{j}^{z}}
    &=\frac{1}{4}\Bigbraket{\left(b_{i\uparrow}^{\dagger}b_{i\uparrow}-b_{i\downarrow}^{\dagger}b_{i\downarrow}\right)\left(b_{j\uparrow}^{\dagger}b_{j\uparrow}-b_{j\downarrow}^{\dagger}b_{j\downarrow}\right)}\nonumber\\
    &\approx\frac{1}{4}\left(|\braket{b_{i\uparrow}b_{j\uparrow}}|^2+|\braket{b_{i\uparrow}^{\dagger}b_{j\uparrow}}|^2+\cdots\right).
\end{align}
In this expression, the Wick decomposition generates pairing channels such as $\braket{b_{i\uparrow}b_{j\uparrow}}$.
Since such pairing channels are present in $\braket{\mathcal{D}_{ij}^x}$ as shown in Eq.~\eqref{eq:SU(2) breaking operators D}, they yield a positive contribution to $\braket{S_{i}^{z}S_{j}^{z}}$ on the $x$ bond.
These FM-like correlations in decoupling I arise because the Wick decomposition is performed in terms of the original bosonic operators, even though the mean-field theory is formulated using the bond operators $\bm{\mathcal{Q}}_{ij}$.
In other words, the pairing channels that appear in decoupling I are not determined self-consistently by SBMFT.
When such pairing channels contribute to the two-point correlator, they generate FM-like contributions that are not determined by the self-consistent bond-operator mean fields, thereby highlighting the advantage of decoupling II.

We next discuss the relationship between our findings and previous studies on the $S=1$ Kitaev model.
In the Schwinger boson approach, an instability toward magnetic ordering is signaled by the vanishing of the bosonic spinon gap, indicating that a QSL phase is stabilized by a finite spinon gap; within this framework, only gapped QSL states occur away from the critical point.
Nevertheless, we find that the bosonic spinon gap for $S=1$ in the self-consistent solution $\omega_{\text{min}}/|K|=0.013$ is vanishingly small, suggesting that the $S=1$ Kitaev model is on the verge of magnetic ordering.
This near-gapless behavior is also reflected in the low-energy spin excitations, which appear in the dynamical structure factors shown in Figs.~\ref{fig:ground_state}(b) and \ref{fig:ground_state}(c).
These features are consistent with tensor-network studies discussing the spin-gap structure of the $S=1$ Kitaev model~\cite{Lee-Kawashima-Kim-2020,Khait-Stavropoulos-2021,Lee-Suzuki-Kim-Kawashima-2021,Chen-2022}. 
Since the minimum spinon excitation energy in our spectrum is $\omega_{\min}/|K|=0.013$, we may estimate the spin gap by identifying it with the lowest two-spinon excitation, yielding $\Delta/|K|\simeq 0.026$. 
This estimate lies within the upper bound on the spin gap, $\Delta/|K|\lesssim 0.04-0.05$, obtained from density-matrix renormalization group calculations for the same model~\cite{Khait-Stavropoulos-2021}. 
On the other hand, other density-matrix renormalization group calculations~\cite{Dong-Sheng-2020} as well as an exact-diagonalization analysis~\cite{Koga-2018} have reported features characteristic of a gapless QSL in the $S=1$ Kitaev model, thereby indicating the need for a more careful investigation of this issue.
As shown in Fig.~\ref{fig:dispersion}(b), the present mean-field ansatz is unstable for $S\gtrsim 1.07$, meaning that the QSL phase is not realized for the $S=3/2$ case within the present Schwinger boson framework.
On the other hand, a previous study using a pf-FRG method found that the QSL phase persists up to $S=3/2$ but is unstable for $S\geq 2$~\cite{Fukui-Kato-Nasu-Motome-2022}.
A possible origin of this discrepancy is the presence of another QSL phase that is not captured by the present mean-field ansatz.
One candidate is a $\pi/2$-flux state, which has been proposed to be more stable for larger $S$ in previous work using SBMFT~\cite{Ralko-Merino-2024}.
As this ansatz explicitly breaks time-reversal symmetry, one must examine whether such a chiral QSL can be realized even in the pure Kitaev Hamiltonian without additional interactions.
Alternatively, our mean-field decoupling of the quartic Schwinger boson interactions may underestimate the bosonic spinon gap, thereby rendering the QSL state unstable.

Finally, we comment on the relationship to previous studies of higher-$S$ Kitaev models based on Majorana-fermion representations~\cite{Ma-2023}.
In such approaches, the bosonic excitations in the $S=1$ Kitaev model are described as a composite object made of two Majorana fermions, which is referred to as a ``giant parton''.
Our Schwinger boson treatment offers a complementary description of these excitations.
Because the present SBMFT addresses quasiparticle excitations of the $S=1$ Kitaev model more directly, it may provide insights into low-energy excitations and the gauge structure that are inaccessible to the Majorana-fermion description alone.
Nevertheless, the relationship between these two descriptions remains unclear and further investigation within a well-controlled mean-field framework is required.

\section{Summary}
\label{sec:Summary}
In summary, we have studied the $S=1$ Kitaev model within the Schwinger boson mean-field framework, where bosonic spinons are introduced as fractional quasiparticles.  
To address the Ising-type anisotropic interactions, we introduced SU(2)-breaking bond operators in addition to the conventional SU(2)-invariant ones.  
Within the present treatment, we found a self-consistent quantum spin-liquid solution.
The resulting spin dynamics exhibits a nonzero but significantly small spinon gap at zero temperature, which gives rise to low-energy continuum structures in the dynamical spin structure factor.
At finite temperatures, this continuum splits into a quasi-elastic component and a higher-energy broad structure.
The temperature evolution of the spectral weight can be understood in terms of the narrowing of the spinon bandwidth, providing insight into the finite-temperature properties of the $S=1$ Kitaev quantum spin liquid.
In evaluating the spin structure factors, we examined two distinct decoupling schemes for the spin correlation functions.
One is the conventional decoupling applied to spinons that has been widely used in previous studies, and the other is the scheme proposed in the present work, which is designed to be consistent with the mean-field ansatz for the bond operators.
We demonstrated that the choice of decoupling scheme significantly affects the resulting spin structure factors.
In particular, our proposed scheme removes the FM-like contributions that are not consistent with the AFM exchange sign structure in the time-reversal-symmetric mean-field ansatz, thereby making this ansatz a useful representative for studying spin dynamics within the Schwinger boson mean-field framework.

Several issues remain for future work.
An important open question is how the bosonic excitations assumed in Schwinger boson theory can be related to the Majorana fermion description employed in a previous study~\cite{Ma-2023}.
Magnetic interactions beyond the Kitaev coupling can be addressed within the present framework by extending the model Hamiltonian, which is crucial for discussing the connection to experimental results on candidate materials for the $S=1$ Kitaev model.
In this context, it will also be important to assess the role of additional degrees of freedom beyond dipolar spin moments that may become relevant in $S=1$ materials, such as single-ion anisotropy, biquadratic interactions, or crystalline-field effects.
While the present work focuses on dipolar spin dynamics within an SU($2$) Schwinger boson framework governed by bilinear exchange interactions, incorporating such effects may require an SU($3$) formulation.
Developing a corresponding SU($3$) Schwinger boson description and applying it to extended $S=1$ Kitaev models therefore constitutes an interesting direction for future work.

\begin{acknowledgments}
  The authors thank A.~Ono, R.~Iwazaki, and S.~Koyama for fruitful discussions. We also thank Y.~Kamiya for valuable advice and fruitful discussions.
  D.S. is grateful to R.~Samajdar, N.~B.~Perkins and C.~D.~Batista for insightful comments and useful advice.
  Parts of the numerical calculations were performed in the supercomputing systems in ISSP, the University of Tokyo.
  This work was supported by Grant-in-Aid for Scientific Research from
  JSPS, KAKENHI Grant Nos.~JP23H01129, JP23H04865, JP24K00563.
  D.S. acknowledges support from GP-Spin at Tohoku University.
\end{acknowledgments}

\appendix

\section{Representation of spin interactions in the Schwinger boson theory}
\label{app:Representation of spin interactions in the Schwinger boson theory}

In this appendix, we derive the Schwinger boson representation of a general two-spin interaction using SU($2$)-breaking bond operators.
We begin by considering the Ising interaction $S_{i}^{\gamma}S_{j}^{\gamma}$.
This interaction can be represented using the following identities:
\begin{align}
    \label{eq:Heisenberg identity}
    \sigma_{\mu\nu}^{x}\sigma_{\rho\lambda}^{x}
    &=\sigma_{\mu\lambda}^{0}\sigma_{\nu\rho}^{0}-\sigma_{\mu\rho}^{z}\sigma_{\nu\lambda}^{z}
    =\sigma_{\mu\lambda}^{x}\sigma_{\nu\rho}^{x}+\sigma_{\mu\rho}^{y}\sigma_{\nu\lambda}^{y},\\
    \sigma_{\mu\nu}^{y}\sigma_{\rho\lambda}^{y}
    &=\sigma_{\mu\lambda}^{0}\sigma_{\nu\rho}^{0}-\sigma_{\mu\rho}^{0}\sigma_{\nu\lambda}^{0}
    =-\sigma_{\mu\lambda}^{y}\sigma_{\nu\rho}^{y}+\sigma_{\mu\rho}^{y}\sigma_{\nu\lambda}^{y},\\
    \sigma_{\mu\nu}^{z}\sigma_{\rho\lambda}^{z}
    &=\sigma_{\mu\lambda}^{0}\sigma_{\nu\rho}^{0}-\sigma_{\mu\rho}^{x}\sigma_{\nu\lambda}^{x}
    =\sigma_{\mu\lambda}^{z}\sigma_{\nu\rho}^{z}+\sigma_{\mu\rho}^{y}\sigma_{\nu\lambda}^{y},
\end{align}
which lead to the following bond-operator representation of the Ising-type interaction:
\begin{align}
    \label{eq:Ising Schwinger boson representation}
    S_{i}^{\gamma}S_{j}^{\gamma}
    &=:\mathcal{B}_{ij}^{\dagger}\mathcal{B}_{ij}:-\mathcal{D}_{ij}^{\gamma\dagger}\mathcal{D}_{ij}^{\gamma}=:\mathcal{C}_{ij}^{\gamma\dagger}\mathcal{C}_{ij}^{\gamma}:-\mathcal{A}_{ij}^{\dagger}\mathcal{A}_{ij}\nonumber\\
    &=\frac{1}{2}\left(:\mathcal{B}_{ij}^{\dagger}\mathcal{B}_{ij}:-\mathcal{D}_{ij}^{\gamma\dagger}\mathcal{D}_{ij}^{\gamma}+:\mathcal{C}_{ij}^{\gamma\dagger}\mathcal{C}_{ij}^{\gamma}:-\mathcal{A}_{ij}^{\dagger}\mathcal{A}_{ij}\right).
\end{align}
In addition, we note that the Heisenberg interaction can be expressed as
\begin{align}
    \label{eq:relation Heisenberg and Ising}
    \bm{S}_{i}\cdot\bm{S}_{j}=\sum_{\gamma=x,y,z}S_{i}^{\gamma}S_{j}^{\gamma}.
\end{align}
Consequently, we obtain the following representation in terms of bond operators:
\begin{align}
    \label{eq:Heisenberg interaction Schwinger boson representation ver2}
    \bm{S}_{i}\cdot\bm{S}_{j}
    =:\mathcal{B}_{ij}^{\dagger}\mathcal{B}_{ij}:-\mathcal{A}_{ij}^{\dagger}\mathcal{A}_{ij}
    =\sum_{\gamma=x,y,z}\left(\mathcal{D}_{ij}^{\gamma\dagger}\mathcal{D}_{ij}^{\gamma}-:\mathcal{C}_{ij}^{\gamma\dagger}\mathcal{C}_{ij}^{\gamma}:\right).
\end{align}

We next consider the off-diagonal spin interaction $S_{i}^{\alpha}S_{j}^{\beta}$ with $\alpha \neq \beta$.  
The Pauli matrices satisfy the following identities:  
\begin{align}  
    \label{eq:off-diagonal identity}  
    \sigma_{\mu\nu}^{x}\sigma_{\rho\lambda}^{y}  
    &=-\sigma_{\mu\lambda}^{x}\sigma_{\nu\rho}^{y}+\sigma_{\mu\rho}^{x}\sigma_{\nu\lambda}^{y}  
    =\sigma_{\mu\lambda}^{y}\sigma_{\nu\rho}^{x}-\sigma_{\mu\rho}^{y}\sigma_{\nu\lambda}^{x}\nonumber\\  
    &=-i\sigma_{\mu\rho}^{z}\sigma_{\nu\lambda}^{0}+i\sigma_{\mu\lambda}^{z}\sigma_{\nu\rho}^{0}=i\sigma_{\mu\rho}^{0}\sigma_{\nu\lambda}^{z}-i\sigma_{\mu\lambda}^{0}\sigma_{\nu\rho}^{z},\\  
    \sigma_{\mu\nu}^{y}\sigma_{\rho\lambda}^{z}  
    &=\sigma_{\mu\lambda}^{y}\sigma_{\nu\rho}^{z}-\sigma_{\mu\rho}^{z}\sigma_{\nu\lambda}^{y}  
    =-\sigma_{\mu\lambda}^{z}\sigma_{\nu\rho}^{y}+\sigma_{\mu\rho}^{y}\sigma_{\nu\lambda}^{z}\nonumber\\  
    &=-i\sigma_{\mu\rho}^{0}\sigma_{\nu\lambda}^{x}+i\sigma_{\mu\lambda}^{x}\sigma_{\nu\rho}^{0}=i\sigma_{\mu\rho}^{x}\sigma_{\nu\lambda}^{0}-i\sigma_{\mu\lambda}^{0}\sigma_{\nu\rho}^{x},\\  
    \sigma_{\mu\nu}^{z}\sigma_{\rho\lambda}^{x}  
    &=\sigma_{\mu\lambda}^{z}\sigma_{\nu\rho}^{x}+i\sigma_{\mu\rho}^{0}\sigma_{\nu\lambda}^{y}  
    =\sigma_{\mu\lambda}^{x}\sigma_{\nu\rho}^{z}+i\sigma_{\mu\rho}^{y}\sigma_{\nu\lambda}^{0}\nonumber\\  
    &=\sigma_{\mu\rho}^{x}\sigma_{\nu\lambda}^{z}+i\sigma_{\mu\lambda}^{y}\sigma_{\nu\rho}^{0}=\sigma_{\mu\rho}^{z}\sigma_{\nu\lambda}^{x}+i\sigma_{\mu\lambda}^{0}\sigma_{\nu\rho}^{y}.  
\end{align}  
Using the above identities, the off-diagonal interaction can be rewritten as  
\begin{align}  
    \label{eq:off-diagonal interaction Schwinger boson representation}  
    &S_{i}^{\alpha}S_{j}^{\beta}\nonumber\\  
    &=\frac{1}{2}\left[:\mathcal{C}_{ij}^{\alpha\dagger}\mathcal{C}_{ij}^{\beta}:+:\mathcal{C}_{ij}^{\beta\dagger}\mathcal{C}_{ij}^{\alpha}:+i\sum_\gamma \epsilon_{\alpha\beta\gamma}\left(\mathcal{D}_{ij}^{\gamma\dagger}\mathcal{A}_{ij}-\mathcal{A}_{ij}^{\dagger}\mathcal{D}_{ij}^{\gamma}\right)\right]\nonumber\\  
    &=-\frac{1}{2}\left[\mathcal{D}_{ij}^{\alpha\dagger}\mathcal{D}_{ij}^{\beta}+\mathcal{D}_{ij}^{\beta\dagger}\mathcal{D}_{ij}^{\alpha}+i\sum_\gamma \epsilon_{\alpha\beta\gamma}\left(:\mathcal{C}_{ij}^{\gamma\dagger}\mathcal{B}_{ij}:-:\mathcal{B}_{ij}^{\dagger}\mathcal{C}_{ij}^{\gamma}:\right)\right],  
\end{align}  
where $\epsilon_{\alpha\beta\gamma}$ denotes the fully antisymmetric Levi-Civita tensor.  
In deriving Eq.~\eqref{eq:off-diagonal interaction Schwinger boson representation}, we impose the Hermiticity condition $\left(S_{i}^{\mu}S_{j}^{\nu}\right)^{\dagger}=S_{i}^{\mu}S_{j}^{\nu}$.

\section{Four-sublattice transformation of the Kitaev model}
\label{app:four-sublattice transformation of the Kitaev model}

\begin{figure}[t]
  \centering
      \includegraphics[width=\columnwidth,clip]{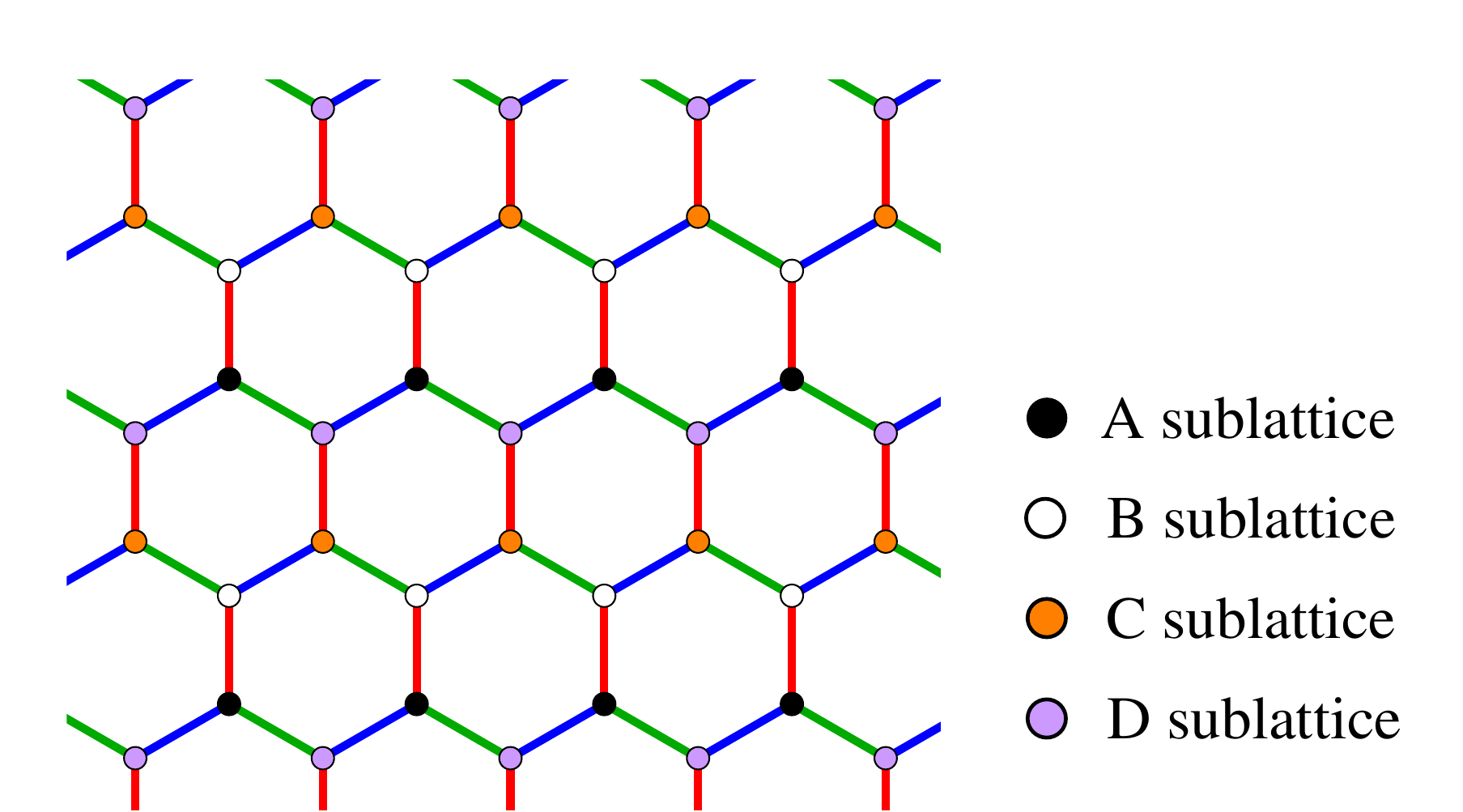}
      \caption{
        Four-sublattice decomposition of the honeycomb lattice used in Appendix~\ref{app:four-sublattice transformation of the Kitaev model}.
        Sites belonging to each sublattice are depicted by colored circles.
      }
      \label{fig:four_sub}
\end{figure}

In this appendix, we demonstrate that the FM and AFM Kitaev Hamiltonians are connected by a unitary transformation corresponding to a sublattice-dependent spin rotation, as introduced in Sec.~\ref{sec:Model}.
Importantly, the four-sublattice transformation is an exact unitary mapping of the spin operators.
As a result, it holds for the spin-$S$ Kitaev Hamiltonian for arbitrary $S$, independent of the Schwinger boson framework.
Therefore, once the AFM Kitaev model is solved within SBMFT, the corresponding FM solution and its dynamical properties can be obtained without additional approximations by applying this transformation.

We consider the honeycomb lattice, which is enlarged into a four-sublattice magnetic unit cell labeled A, B, C, and D, as shown in Fig.~\ref{fig:four_sub}.
The AFM Kitaev Hamiltonian is given by
\begin{align}
\label{eq:AFM Kitaev}
    \mathcal{H}_{\text{AFM}}=K\sum_{\braket{i,j}_{\gamma}}S_{i}^{\gamma}S_{j}^{\gamma},
\end{align}
where $K$ denotes a positive constant.
To map Eq.~\eqref{eq:AFM Kitaev} onto its FM counterpart, we apply the sublattice-dependent unitary transformation to the spin located at site $i$ belonging to the sublattice $\Lambda$ as
\begin{align}
\label{eq:spin transformation}
    \widetilde{\bm{S}}_{i}=U_{\Lambda}\,\bm{S}_{i}\,U_{\Lambda}^{\dagger},
\end{align}
where $\Lambda\in\{\mathrm{A},\mathrm{B},\mathrm{C},\mathrm{D}\}$, and $U_{\Lambda}\in\mathrm{SU}(2)$ is defined as
\begin{align}
    \label{eq:four-sublattice-rotation-operators}
    U_{\text{A}}=1,\:\:U_{\text{B}}=e^{-i\pi S^{x}},\:\:U_{\text{C}}=e^{-i\pi S^{y}},\:\:U_{\text{D}}=e^{-i\pi S^{z}}.
\end{align}
These $\pi$ rotations leave $S_{i}^{\gamma}$ invariant along the rotation axis while flipping the sign of the two orthogonal components.
Substituting Eq.~\eqref{eq:spin transformation} into Eq.~\eqref{eq:AFM Kitaev} yields
\begin{align}
    \label{eq:afm-to-fm-kitaev-mapping}
    \mathcal{H}_{\text{AFM}}\longrightarrow\widetilde{\mathcal{H}}_{\text{AFM}}=K\sum_{\braket{i,j}_{\gamma}}\widetilde{S}_{i}^{\gamma}\widetilde{S}_{j}^{\gamma}=-K\sum_{\braket{i,j}_{\gamma}}S_{i}^{\gamma}S_{j}^{\gamma}\equiv\mathcal{H}_{\text{FM}},
\end{align}
which corresponds to the FM Kitaev Hamiltonian with coupling $-K<0$.
Therefore, the AFM and FM models are connected by a unitary transformation: all observables can be mapped from one to the other by the sublattice rotation in Eq.~\eqref{eq:spin transformation}.

\section{Spin structure factors in other flux sectors}
\label{app:spin-structure-factors-in-other-flux-sectors}

To label the flux sector of a pairing ansatz, we introduce Wilson-loop phases on an oriented elementary hexagonal plaquette $i\to j\to k\to l\to m\to n\to i$.
The successive bond types are chosen as $x,y,z,x,y,z$.
For the spin-singlet pairing channel, we define $W_s \equiv \braket{\mathcal{A}_{ij}}\left(-\braket{\mathcal{A}_{jk}}^{*}\right)\braket{\mathcal{A}_{kl}}\left(-\braket{\mathcal{A}_{lm}}^{*}\right)\braket{\mathcal{A}_{mn}}\left(-\braket{\mathcal{A}_{ni}}^{*}\right)$ and its phase by
\begin{align}
    e^{i\phi_{s}}
    &=
    \frac{W_s}{|W_s|}.
    \label{eq:singlet_flux}
\end{align}
Similarly, for the spin-triplet pairing channel, we define $W_t \equiv \braket{\mathcal{D}_{ij}^{x}}\braket{\mathcal{D}_{jk}^{y}}^{*}\braket{\mathcal{D}_{kl}^{z}}\braket{\mathcal{D}_{lm}^{x}}^{*}\braket{\mathcal{D}_{mn}^{y}}\braket{\mathcal{D}_{ni}^{z}}^{*}$ and its phase by
\begin{align}
    e^{i\phi_{t}}
    &=
    \frac{W_t}{|W_t|}.
    \label{eq:triplet_flux}
\end{align}
Here, $\phi_s$ and $\phi_t$ are defined modulo $2\pi$ when the corresponding Wilson-loop products $W_s$ and $W_t$ are nonzero, respectively.
The minus signs multiplying the complex-conjugated factors are those of the standard anomalous-pairing Wilson loop used in Refs.~\cite{Messio-Bieri-2017,Ralko-Merino-2024}.

\begin{figure*}[t]
  \centering
      \includegraphics[width=2\columnwidth,clip]{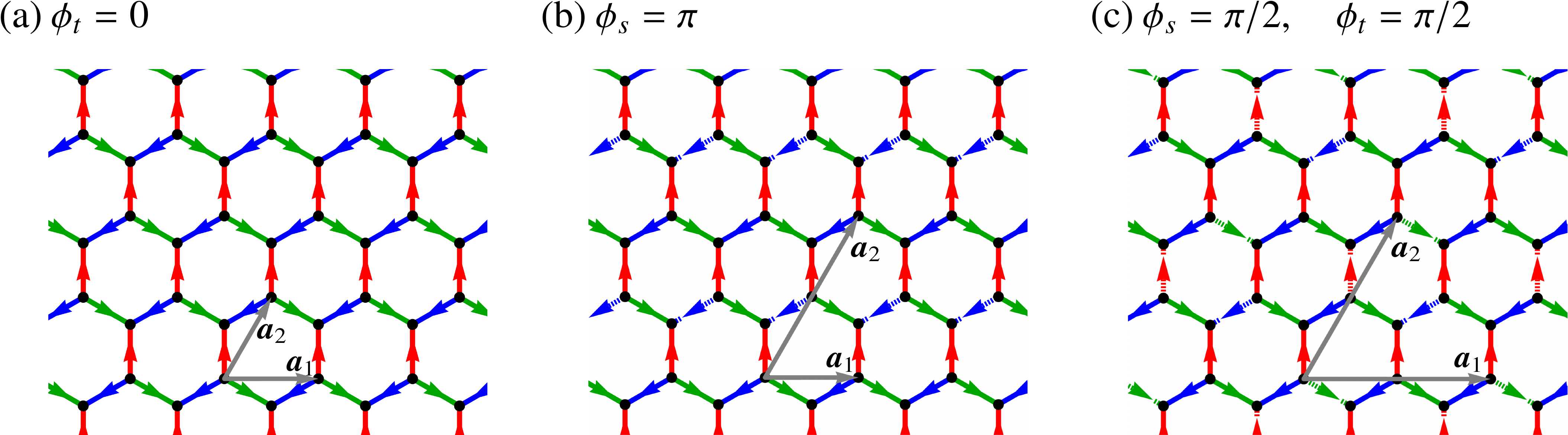}
      \caption{
      Mean-field phase patterns used for the flux sectors considered in this work.
      Arrows specify the direction from site $i$ to site $j$ used to define the mean-field amplitudes.
      In panel (b), the dashed bonds carry an additional factor $-1$ relative to the other bonds.
      In panel (c), the dashed bonds carry an additional factor $i$.
      The nonzero mean field is $\braket{\mathcal{A}_{ij}}$ for the singlet sectors and $\braket{\mathcal{D}_{ij}^{\gamma}}$ on a $\gamma$ bond for the triplet sectors.
      }
      \label{fig:other_flux_mean_field_patterns}
\end{figure*}

\begin{figure*}[t]
  \centering
      \includegraphics[width=2\columnwidth,clip]{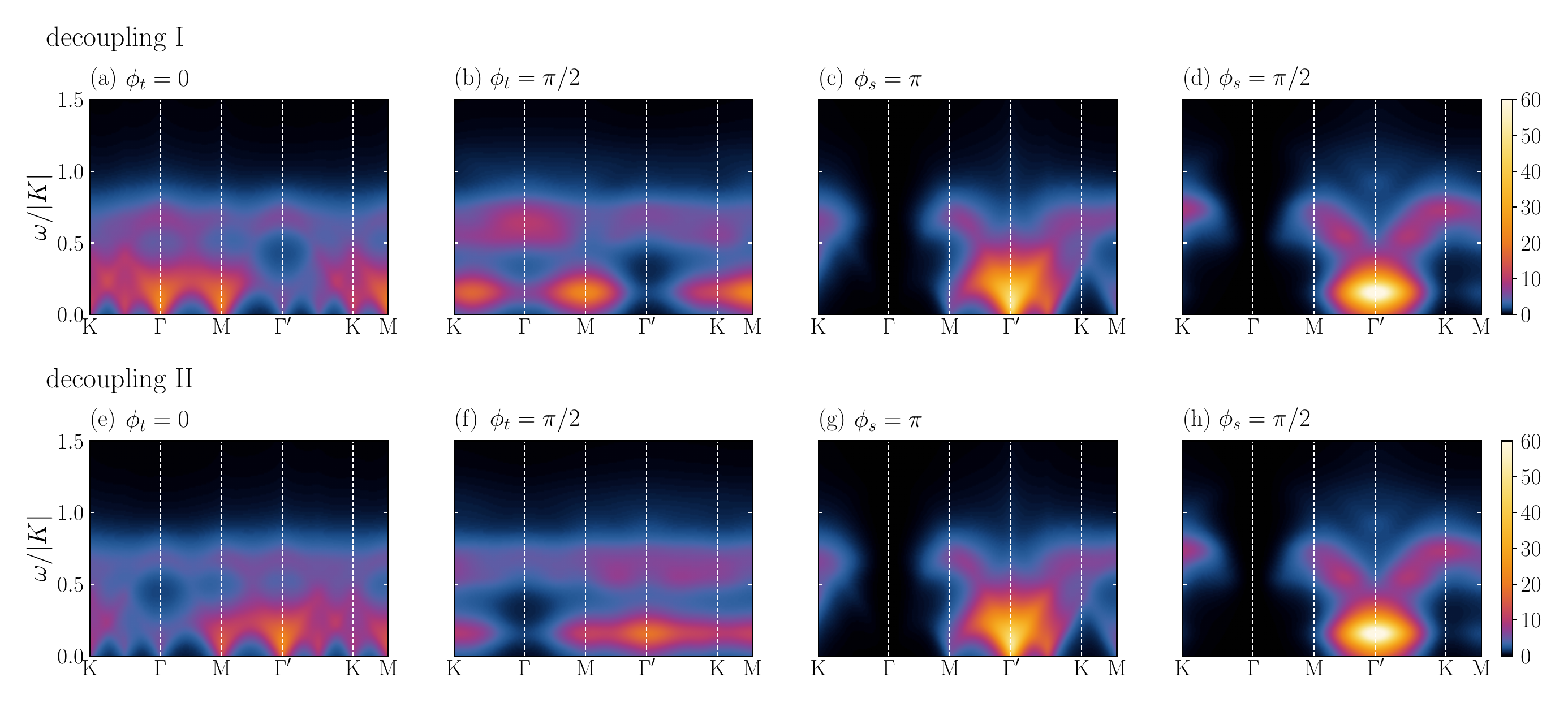}
      \caption{
      Zero-temperature dynamical spin structure factors for different flux sectors of the $S=1$ AFM Kitaev model.
      The upper and lower rows show the results obtained using decoupling I and decoupling II, respectively.
      The four columns correspond to $\phi_t=0$, $\phi_t=\pi/2$, $\phi_s=\pi$, and $\phi_s=\pi/2$.
      }
      \label{fig:other_flux_dssf}
\end{figure*}

\begin{figure*}[t]
  \centering
      \includegraphics[width=2\columnwidth,clip]{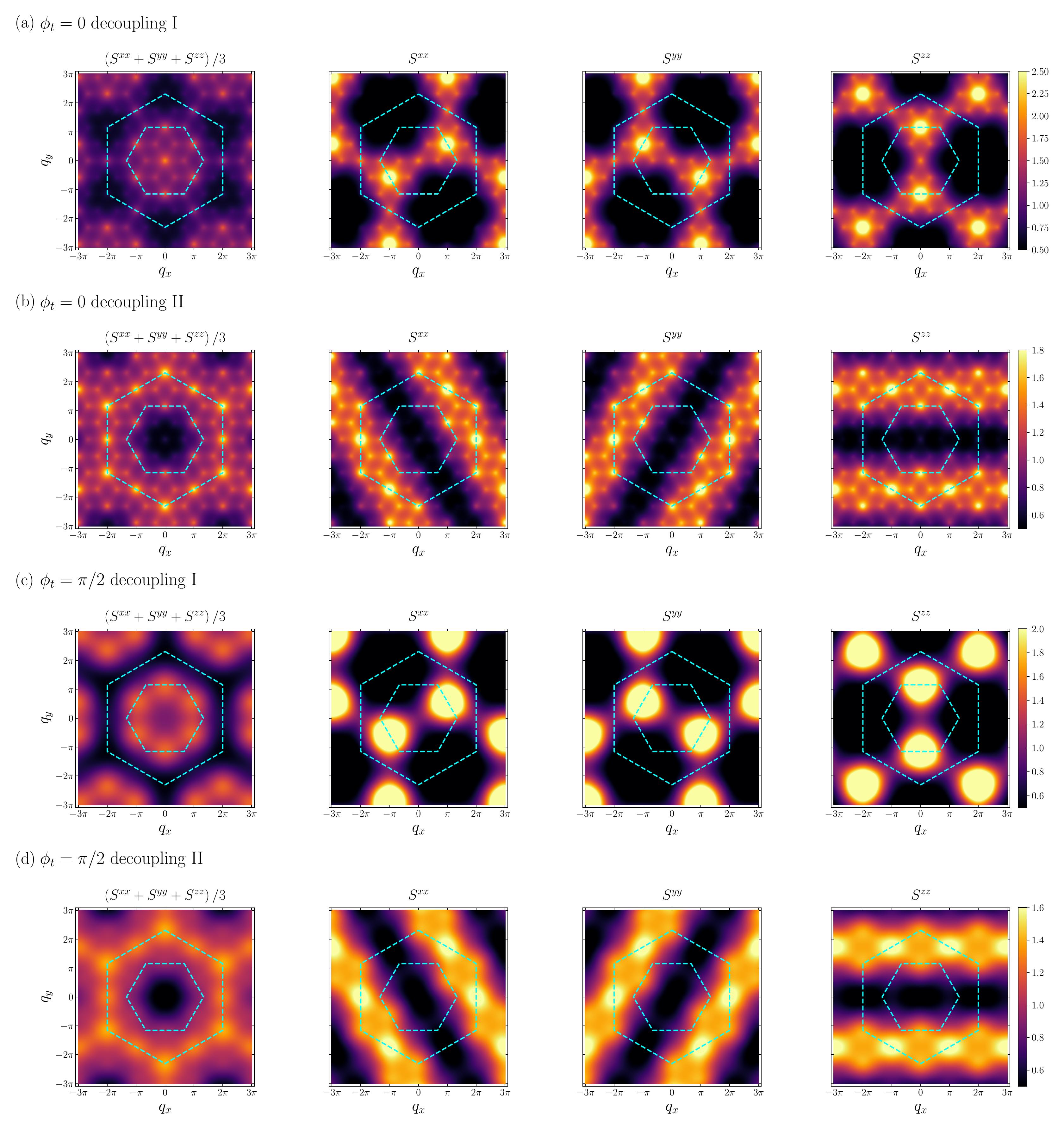}
      \caption{
      Component-resolved equal-time spin structure factors for the spin-triplet flux sectors of the $S=1$ AFM Kitaev model.
      Panels (a) and (b) show the $\phi_t=0$ state evaluated with decoupling I and decoupling II, respectively, while panels (c) and (d) show the $\phi_t=\pi/2$ state.
      In each row, the first panel shows the trace $(S^{xx}+S^{yy}+S^{zz})/3$, and the remaining panels show $S^{xx}$, $S^{yy}$, and $S^{zz}$.
      }
      \label{fig:other_triplet_ansatz_SSF}
\end{figure*}

\begin{figure*}[t]
  \centering
      \includegraphics[width=2\columnwidth,clip]{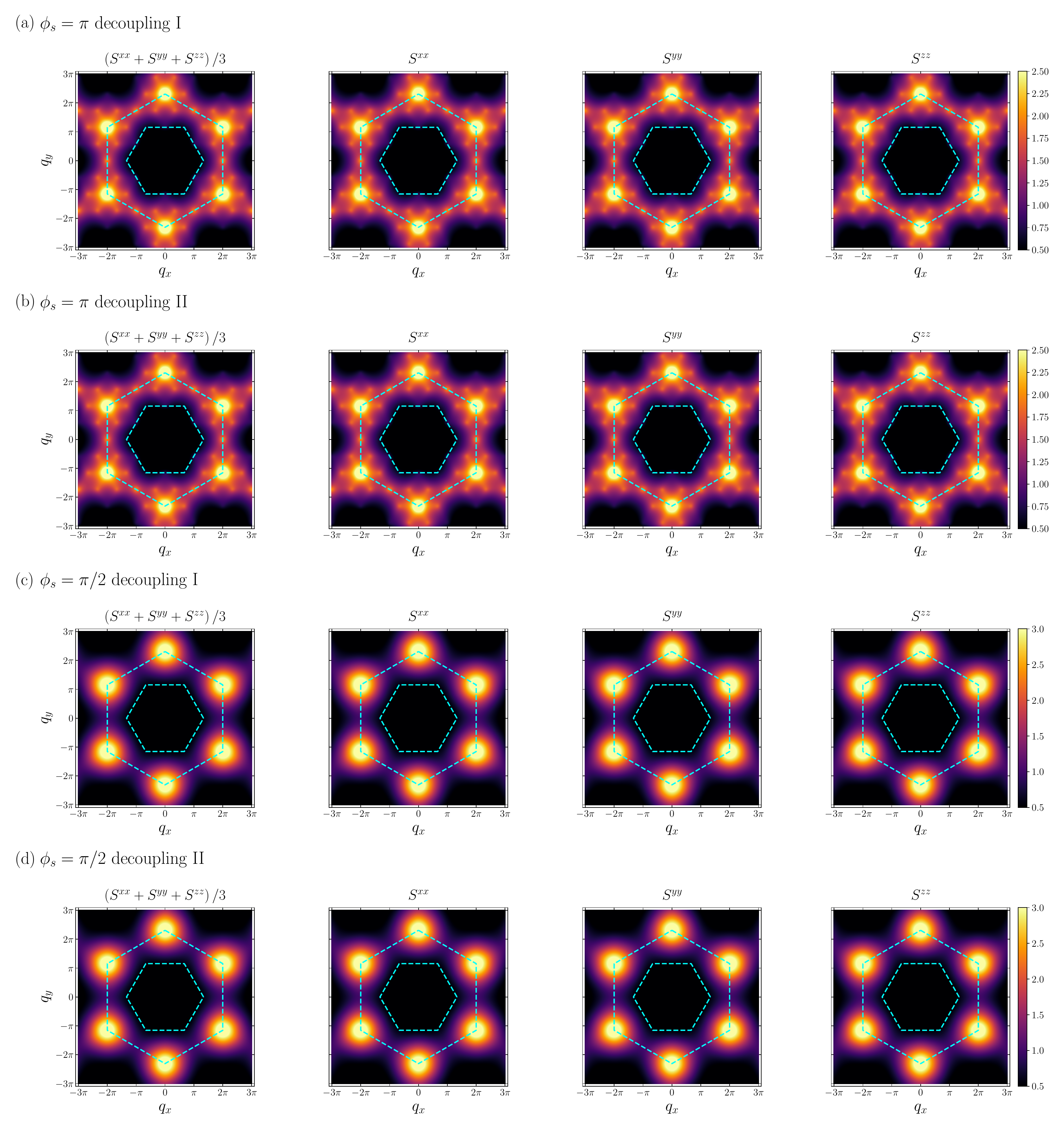}
      \caption{
      Component-resolved equal-time spin structure factors for the spin-singlet flux sectors of the $S=1$ AFM Kitaev model.
      Panels (a) and (b) show the $\phi_s=\pi$ state evaluated with decoupling I and decoupling II, respectively, while panels (c) and (d) show the $\phi_s=\pi/2$ state.
      In each row, the first panel shows the trace $(S^{xx}+S^{yy}+S^{zz})/3$, and the remaining panels show $S^{xx}$, $S^{yy}$, and $S^{zz}$.
      For the spin-singlet ansatze, the three spin components exhibit identical momentum dependence, reflecting the SU($2$)-invariant structure of the singlet mean-field channel.
      }
      \label{fig:other_singlet_ansatz_SSF}
\end{figure*}

In the main text, we focus on the zero-flux spin-triplet ansatz, $\phi_t=0$, as a representative time-reversal-symmetric spin-triplet Kitaev spin-liquid state.
For completeness, we here present the zero-temperature spin dynamics for the other spin-liquid flux sectors considered in this work, namely $\phi_t=\pi/2$, $\phi_s=\pi$, and $\phi_s=\pi/2$.
The zero-flux triplet sector $\phi_t=0$ is also shown in Fig.~\ref{fig:other_flux_dssf} as a reference.

Figure~\ref{fig:other_flux_mean_field_patterns} summarizes the phase conventions used for these flux sectors.
For the zero-flux spin-triplet sector shown in Fig.~\ref{fig:other_flux_mean_field_patterns} (a), the mean-field amplitude is taken as
\begin{align}
    \braket{\mathcal{D}_{ij}^{\gamma}}
    =
    D
    \qquad
    \text{on a $\gamma$ bond}.
    \label{eq:appendix-c-zero-flux-triplet-pattern}
\end{align}
This is the ansatz used in the main text.
For the $\pi$-flux spin-singlet sector shown in Fig.~\ref{fig:other_flux_mean_field_patterns} (b), the nonzero mean-field amplitude is assigned as
\begin{align}
    \braket{\mathcal{A}_{ij}}
    =
    \eta_{ij}^{(\pi)} A,
    \qquad
    \eta_{ij}^{(\pi)}
    =
    \begin{cases}
    -1, & \text{on dashed bonds},\\
    +1, & \text{otherwise}.
    \end{cases}
    \label{eq:appendix-c-pi-flux-singlet-pattern}
\end{align}
Here, $A$ is determined self-consistently together with the Lagrange multiplier.
The arrow direction in Fig.~\ref{fig:other_flux_mean_field_patterns} specifies the ordered pair $(i,j)$ used in $\braket{\mathcal{A}_{ij}}$.
For the chiral $\pi/2$-flux sectors shown in Fig.~\ref{fig:other_flux_mean_field_patterns} (c), the phase factor is
\begin{align}
    \eta_{ij}^{(\pi/2)}
    =
    \begin{cases}
    i, & \text{on dashed bonds},\\
    1, & \text{otherwise}.
    \end{cases}
    \label{eq:appendix-c-pi-over-two-phase-factor}
\end{align}
For the spin-singlet chiral sector, the nonzero mean field is
\begin{align}
    \braket{\mathcal{A}_{ij}}
    =
    \eta_{ij}^{(\pi/2)} A,
    \label{eq:appendix-c-pi-over-two-singlet-pattern}
\end{align}
whereas for the spin-triplet chiral sector, the nonzero mean field is
\begin{align}
    \braket{\mathcal{D}_{ij}^{\gamma}}
    =
    \eta_{ij}^{(\pi/2)} D
    \qquad
    \text{on a $\gamma$ bond}.
    \label{eq:appendix-c-pi-over-two-triplet-pattern}
\end{align}
These conventions follow the flux patterns used in Ref.~\cite{Ralko-Merino-2024}.

Figures~\ref{fig:other_flux_dssf}, \ref{fig:other_triplet_ansatz_SSF}, and \ref{fig:other_singlet_ansatz_SSF} show the zero-temperature dynamical and component-resolved equal-time spin structure factors calculated for the four flux sectors.
For the spin-triplet ansatze, $\phi_t=0$ and $\phi_{t}=\pi/2$, the result depends strongly on the decoupling scheme.
This confirms that the choice of decoupling scheme is important when the mean-field Hamiltonian contains the SU($2$)-breaking triplet channel $\braket{\mathcal{D}_{ij}^{\gamma}}$.
By contrast, for the spin-singlet ansatze, $\phi_{s}=\pi$ and $\phi_{s}=\pi/2$, the spectra obtained using decoupling I and decoupling II are essentially identical.
This is because these spin-singlet ansatze contain only the SU($2$)-invariant singlet-pairing channel $\braket{\mathcal{A}_{ij}}$ as the finite intersite mean field.
As shown in Fig.~\ref{fig:other_singlet_ansatz_SSF}, the spin components therefore satisfy $S^{xx}(\bm{q})=S^{yy}(\bm{q})=S^{zz}(\bm{q})$, and decoupling II, which respects the bond-operator truncation used to define the ansatz, gives the same result as decoupling I for these spin-singlet ansatze.

\bibliography{./refs}

\end{document}